\documentclass[apj]{emulateapj}
\journalinfo{\textsc{The Astrophysical Journal}}
\slugcomment{Received: March 14, 2018; Accepted: May 1, 2018}
\usepackage{url}

\usepackage{apjfonts}
\usepackage{lineno}
\usepackage{natbib}
\usepackage{graphicx}

\usepackage{array,etoolbox}
\preto\tabular{\setcounter{magicrownumbers}{0}}
\newcounter{magicrownumbers}

\newcommand{\cjaa}{{\it Chinese Journal of Asronomy \& Astrophysics}}

\shorttitle{magnetic non-potentiality in confined and eruptive flares}
\shortauthors{Vasantharaju et al}
\begin{document}
\title{Statistical study of magnetic non-potential measures in confined and eruptive flares}
\author{N.~Vasantharaju$^1$, P.~Vemareddy$^2$, B.~Ravindra$^2$, and V.~H.~Doddamani$^1$}
\email{vrajuap@gmail.com}
\affil{$^1$Department of Physics, Bangalore University, Bengaluru-560 056, India}
\affil{$^2$Indian Institute of Astrophysics, Koramangala, Bengaluru-560 034, India}

\begin{abstract}
Using the HMI/SDO vector magnetic field observations, we studied the relation of degree of magnetic non-potentiality with the observed flare/CME in active regions. From a sample of 77 flare/CME cases,  we found a general relation that degree of non-potentiality is positively correlated with the flare strength and the associated CME speeds. Since the magnetic flux in the flare-ribbon area is more related to the reconnection, we trace the strong gradient polarity inversion line (SGPIL), Schrijver's R value manually along the flare-ribbon extent. Manually detected SGPIL length and R values show higher correlation with the flare strength and CME speed than the automatically traced values without flare-ribbon information. It highlights the difficulty of predicting the flare strength and CME speed a priori from the pre-flare magnetograms used in flare prediction models. Although the total, potential magnetic energy proxies show weak positive correlation, the decrease in free energy exhibits higher correlation (0.56) with the flare strength and CME speed. Moreover, the eruptive flares have threshold of SGPIL length (31Mm), R value ($1.6\times10^{19}$Mx), free-energy decrease ($2\times10^{31}$erg) compared to confined ones. In 90\% eruptive flares, the decay-index curve is steeper reaching $n_{crit}=1.5$ within 42Mm, whereas it is beyond 42Mm in $>70$\% confined flares. While indicating the improved statistics in the predictive capability of the AR eruptive behaviour with the flare-ribbon information, our study provides threshold magnetic properties for a flare to be eruptive. 
\end{abstract}


%
%
\section{Introduction}
\label{Intro}

The energetic events like flares and coronal mass ejections (CMEs) generally occur in magnetically concentrated locations called active regions (ARs). It is believed that these events are physically related to magnetic complexity and non-potentiality of ARs.  However, the question of how the magnetic complexity and non-potentiality trigger the flares and CMEs remains to be unknown. For this reason, solar flare forecasting models mainly depend on statistical relationship between the flare production and the non-potential magnetic parameters characterizing the AR size, strength, morphology, magnetic topology etc. Conventionally, the magnetic complexity and non-potentiality are described with parameters such as the magnetic shear \citep{Hagyard1986,Wang1994}, the horizontal gradient of longitudinal magnetic field \citep{Zirin1993a, Tian2002}, the electric current \citep{Leka1993, WangT1994}, twist parameter $\alpha$ \citep{Pevtsov1994,Hagino2004, Tiwari2009}, magnetic free energy \citep{Metcalf2005a}, current helicity \citep{Abramenko1996,ZhangHQ1999} etc. Although individual cases have identified a direct role of these non-potential parameters in the observed activity \citep{Vemareddy2012}, their relationship is not quite strong enough in statistically significant sample of flare/CME and hence for the prediction of space-weather.


Past observational studies have explored the connection between photospheric magnetic fields and solar flares, supporting the claim that solar flares are driven by the non-potentiality of magnetic fields in ARs \citep{Leka1993,WangT1994,WangJ1996,Tian2002,Abramenko2005}. Several observations have revealed that solar flares often occur near polarity inversion lines (PILs) with high gradient of longitudinal magnetic fields and/or strong shear of transverse components \citep{Hagyard1990,Wang1994,Falconer1997,Kosovichev2001,Jing2006,Schrijver2007}. Because of this fact, the magnetic shear, magnetic gradient and the vertical electric currents in the photosphere are the most commonly used measures of the magnetic non-potentiality. 

Both gradient and shear have been employed as parameters in solar activity forecast models. For example, in a sample of 17 vector magnetograms, \citet{Falconer2001} and \citet{Falconer2003} measured the lengths of strong-sheared and strong-gradient magnetic neutral line segments, respectively. Their study found that strong-sheared, strong-gradient PIL segments are strongly correlated with each other, suggesting them to be prospective predictors of the CME productivity of ARs. \citet{Leka2003a,Leka2003b} investigated the magnitudes and temporal variations of several photospheric magnetic parameters in three ARs. They demonstrated that individually these parameters have little ability to differentiate between flare-producing and flare-quiet regions, but in certain combinations, two populations may be distinguished. \citet{Song2006} showed that the strong-gradient polarity inversion line length (SGPIL) is a viable tool to locate source regions of either CMEs or flares. The definitive flare/CME prediction ability by measuring SGPIL is about 75\% (55 out of 73 events). Based on a study of 298 ARs, \citet{Georgoulis2007} argued the connected magnetic field as the robust metric to distinguish X- and M-flaring regions from non-flaring ones. A further study by \citet{Song2009} proved that SGPIL is the most promising parameter in determining the solar flares, if only one parameter had to choose. In combination with PIL length, total unsigned flux and the effective separation, \citet{Mason2010} introduced the gradient-weighted PIL length as a characteristic for solar flare forecast, however the skill score test still indicates it is not a reliable parameter for real-time predictor of flares. However, \citet{Sadykov2017} showed the importance of PIL characteristics in the solar flare forecasts.

Knowledge of the amount of magnetic free energy and its temporal variation associated with flares/CMEs is important to understand the energy storage and release processes in ARs. The deviation from potential energy is taken as a proxy for magnetic free energy. \citet{Leka2007} considered the total excess energy as one of the best performing parameters. \citet{Emslie2012} demonstrated for a sample of 38 flares that the total magnetic free energy was sufficient to explain the flare energy release including CMEs, energetic particles, and hot plasma emission and dynamics. Studies on the correlation between magnetic free energy estimated from three-dimensional non-linear force-free fields (3D NLFFF) and flare index, confirm the physical link between them \citep{Jing2010}. \citet{Su2014} showed that in flaring ARs, the 3D NLFFF magnetic free energy and the magnetic free energy obtained from photospheric magnetic fields have almost equal predictability for flares. 

Motivated by the above studies, we further investigated the relationship between non-potentiality of magnetic field in statistically significant source ARs and the observed flare/CME productivity using vector magnetograms. The central interest of this study seeks to address the question that the degree of non-potentiality has any correspondence with the magnitude of the flares and the speed of the CMEs. Both flares and CMEs are magnetically driven phenomenon of stored magnetic energy configuration and it is of great importance to distinguish the AR conditions that produce flares and that lead to a flare become eruptive successfully ejecting material as a CME. Note that coronal rain is also a kind of eruption but fail to eject material. Seeking such a link of AR non-potential parameters to flares in the begining and then to CMEs involves a careful manual inspection of which CME comes from which AR and the availability of vector magnetic field measurements within $\pm40^\circ$ of disk center. Therefore, our study is limited in sample compared to several studies based on line-of-sight magnetograms as stated above. In section~\ref{ObsData}, we gave the brief description about the observational data, and the procedures used to calculate the several magnetic field parameters.  The results of non-potential measures and their relation to flares and CMEs are discussed in section~\ref{res}. The summary with discussion is presented in Section~\ref{summ}.

\section{Observational Data and Analysis Procedure}
\label{ObsData}
The required vector magnetic field observations are obtained from the Helioseismic and Magnetic Imager (HMI; \citet{Schou2012}) aboard Solar Dynamics Observatory (SDO). For ready use in AR analysis, HMI provides processed space-weather HMI active region patches (SHARPs) at 720s cadence. This data product contains field components $(B_r, B_\theta, B_\phi)$ in spherical coordinate system after the heliographic CEA projection \citep{Calabretta2002} centered on the AR patch. These components are equivalent to $(B_z,-B_y, B_x)$ in cartesian system. More information on the HMI data pipeline and the field transformation is described in \citet{Hoeksema2014,Sun2013}. The data on CME linear speed, flare initiation and end timings were obtained from relevant websites ({\url{https://cdaw.gsfc.nasa.gov/CME\_list/}} \citep{Gopalswamy2009},
		{\url{https://www.spaceweatherlive.com/en/archive,
		http://xrt.cfa.harvard.edu/flare\_catalog/all.html.}}).
For a statistically significant inference, we considered 77 flare events during the period from February 2011 to July 2016 and their source ARs are located within 40$^{\circ}$ of the central meridian to minimize the projection effects on the calculation of magnetic properties are shown in Figure ~\ref{Fig_eve_pos}. We also excluded the flares from inter-ARs \citep{Toriumi2017} and/or not associated with polarity inversion line. Thus, our sample constitutes 14 X-class, 42 M-class and 21 C-class flares. Among the sample, 38 flares are associated with CMEs visible at least in LASCO C2 field-of-view. The procedures for the derivation of various magnetic parameters are described as follows.

\subsection{Strong gradient and Strong sheared PIL length}
PILs mark the separation between  positive and negative magnetic flux patches in the photosphere of the ARs.  Solar flares generally occur in the strong magnetic regions with strong gradient polarity inversion lines, or in complex polarity patterns. In the past, several researchers measured the PIL length and studied its relationship with flare productivity \citep{Falconer2003,Bokenkamp2007,Schrijver2007,Jing2006,Song2006,Mason2010}. \citet{Bokenkamp2007} developed a three-step iterative algorithm to measure the gradient-weighted polarity inversion line lengths, originally based on the \citet{Falconer2003} method. In the first step, zero gauss contours are identified in the strongly smoothed line of sight magnetic field( $B_{los}$) map and  the vector magnetic field map is calculated from the smoothed $B_{los}$ using linear force-free field model \citep{Alissandrakis1981}. From the output of this step, PIL segments are identified as zero Gauss contours with specific thresholds of simulated potential transverse field and horizontal gradient of $B_{los}$. In the second step, the above process is repeated for the less smoothed $B_{los}$ image and the output PIL segments are identified by comparing it with PIL segments obtained with this step and the previous one. Finally, in the third step, the same process is repeated for the unsmoothed $B_{los}$ and the gradient weighted PIL length is obtained by comparing the segment outputs from this step and the previous step. Here in our study, we employed a single step algorithm to measure the Strong Gradient polarity inversion line (SGPIL) and Strong shear polarity inversion line (SSPIL) lengths based on Bokenkamp's algorithm. These PILs are defined in the following paragraphs. 

Non-potential nature of magnetic field is characterized by magnetic shear as the difference between directions of observed transverse fields ($\mathbf{B}_o$) and potential transverse fields ($\mathbf{B}_p$) and is given by,  $\theta= cos^{-1}(\mathbf{B}_o \cdot \mathbf{B}_p / |\mathbf{B}_o│ \mathbf{B}_p|)$  \citep{Ambastha1993,Wang1994}. The PIL segment on which the observed transverse field has strength above the threshold value of 150G and magnetic shear angle has exceeding value of 45$^{\circ}$ is marked as SSPIL \citep{Falconer2003,Vemareddy2015}.  We measured the SSPILs automatically ($SSPIL_A$) and also manually ($SSPIL_M$). Our algorithm (automatic) identifies multiple strong shear PILs in an AR and sum of these PIL lengths gives $SSPIL_A$. In this procedure, we smoothed  the vertical magnetic field ($B_r$)  image to a smoothing factor of 8 pixels(4 arcseconds) and identified the zero gauss contour. Then potential magnetic field  is calculated from the strongly smoothed magnetogram and thereby shear map is generated from the computed potential transverse field and the observed transverse field. On applying the thresholds of strong observed transverse field, $B_t$ ($>300$G) and strong  shear angle ($>45^{\circ}$) to contour segments, strong field and strong shear PILs are identified. The summation of these PILs lengths gives $SSPIL_A$. 

In the case of vector magnetograms are not available, the shear information alternatively obtained by horizontal gradients in line-of-sight magnetograms where the SGPIL serves the purpose of SSPIL \citep{Falconer2003}. In this procedure (automatic), we identified the zero gauss contours on smoothed magnetogram. Then potential magnetic field and vertical magnetic field gradient maps were generated by the strongly smoothed magnetogram. Using a threshold of potential transverse field ( $>300$G) and high- vertical magnetic field gradient ($>50$ G/Mm) to the zero Gauss contour segments, $SGPIL_A$ is determined. 

To illustrate this procedure, we display AR NOAA 11429 in Figure~\ref{Fig_PIL_FL_rib}. The panel \ref{Fig_PIL_FL_rib}(a) shows flare ribbons in the AIA 1600\AA~snapshot taken during GOES soft X-ray peak time (2012-03-07 00:24 UT) of flare X5.4 from AR 11429. The brightening region indicates the flare occurring site. These flare ribbons generally trace SGPIL/SSPIL with some departures as displayed in overplotted Bz map (Figure~\ref{Fig_PIL_FL_rib}). In panel~\ref{Fig_PIL_FL_rib}(c), we traced automatically the \textbf{$SGPIL_A$} marked by blue curve on gradient map of Bz. This difference slightly increases depending on the complexity of ARs. It can be noticed that both $SGPIL_A$ and {$SSPIL_A$ traces the same segments of PIL and measures a length with slight difference. A similar trace is obtained with \textbf{$SSPIL_A$} as shown in shear map of panel~\ref{Fig_PIL_FL_rib}(d). The $SSPIL_A$ also follows the flare brightening region to a large extent with differences at the middle. In panel~\ref{Fig_PIL_FL_rib}(e), we overplotted $SGPIL_A$ (blue curve) on Bz map and the $SSPIL_A$ (blue) is overlayed on vector magnetogram in panel~\ref{Fig_PIL_FL_rib}(f).

Owing to differences in automatically traced SGPIL/SSPILs with the actual extent of flare ribbons during peak time, we undertook an experiment by following these PIL segments manually with the flare ribbon informations. Different from automated procedure, the manual tracing is biased by the flare brightening area which overlaps with the SSPIL/SGPIL.  In this way, we extract only the strong shear/gradient PIL segment which majorly contributes to the flare brightening. The underlying idea to invoke the information of flare brightening is that the flux along the extent of flare ribbon contributes to strength of the flare but not all PIL segments with strong shear/gradient contributes to the flare intensity. We refer the total length of all these manually traced segments as $SSPIL_M$ and $SGPIL_M$. In future, this manual method can be easily automated (for few cases) as well by applying brightness threshold to AIA 1600~\AA~images and use as mask for HMI magnetograms. However, we are constrained to manually trace these PILs for the reasons of picking peak phase of the flare and verifying the ribbons extent along the PIL instead on either side. This procedure is presently subjective to human errors and may not reproduce the same values.
\begin{figure}
	\centering
	\includegraphics[width=.49\textwidth,clip=]{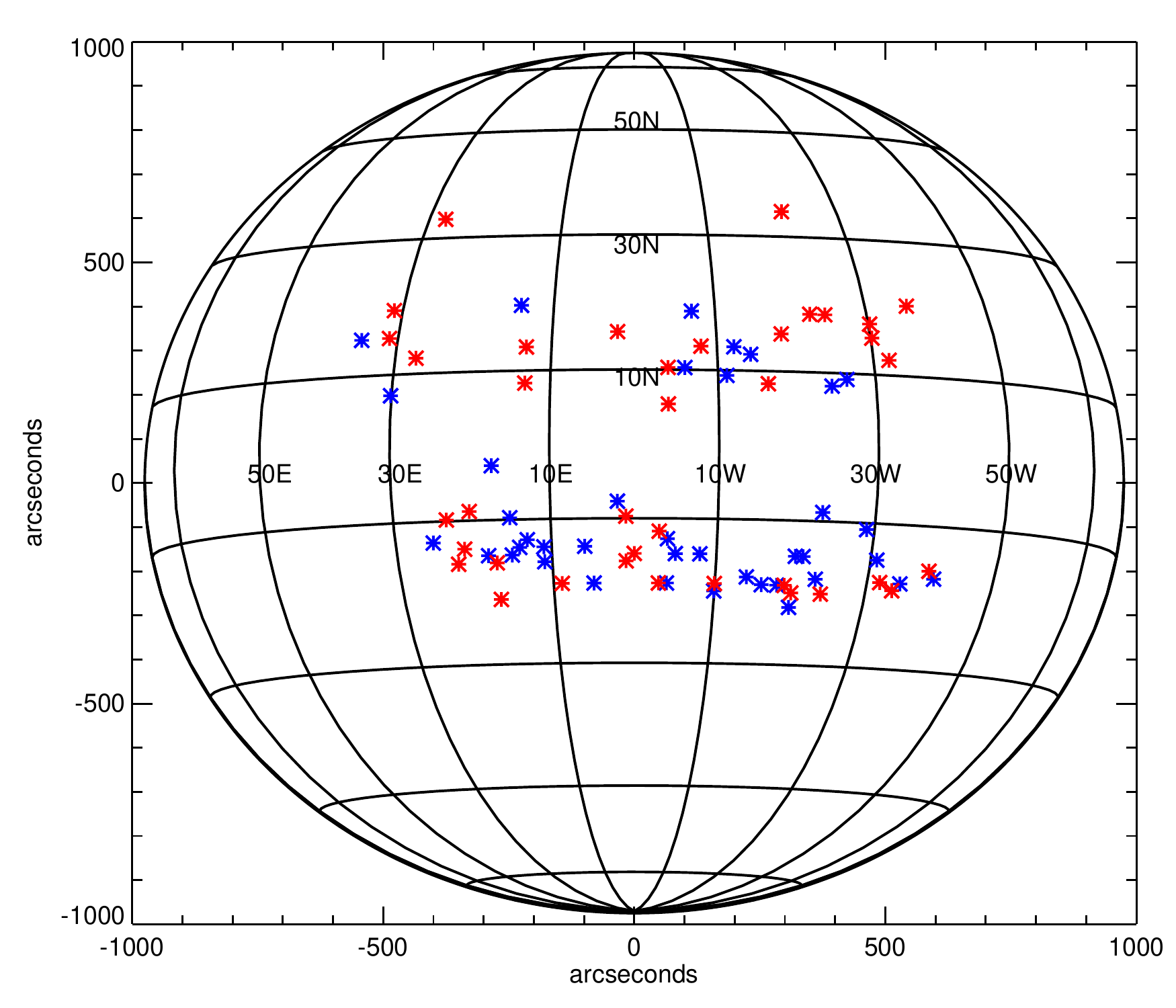}
	\caption{Heliographic locations of all 77 events used in the study. Red (blue) asterisk symbols represent eruptive (confined) flares.}
	\label{Fig_eve_pos}
\end{figure}

\begin{figure*}
	\centering
	\includegraphics[width=.9\textwidth,clip=]{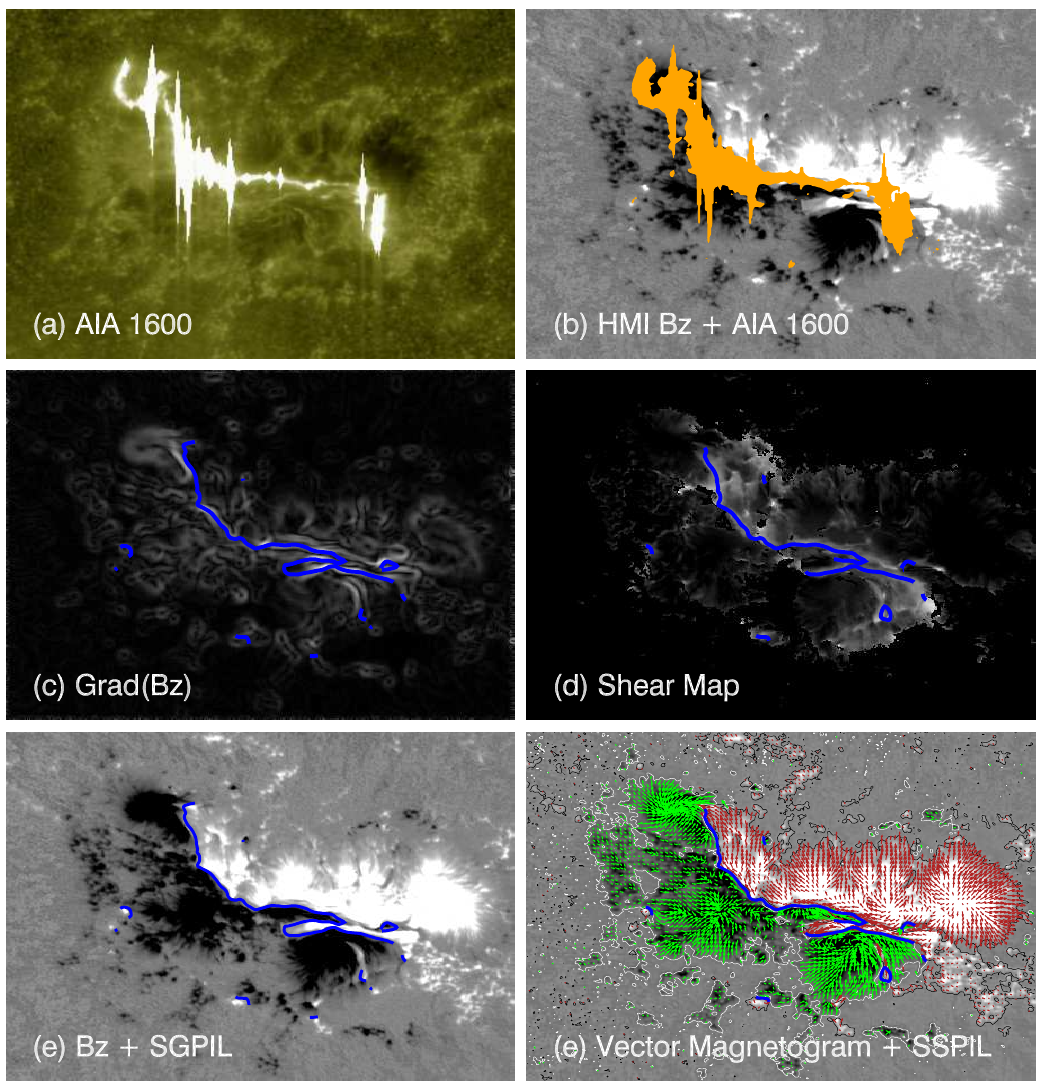}
	\caption{Illustration of tracing $SSPIL_A$ and $SGPIL_A$ in AR 11429 on March 7, 2012. (a) AIA 1600\AA ~image with flare brightening corresponding to peak time of X5.4 flare, b) HMI photospheric magnetogram Bz with filled contours of the AIA 1600\AA~flare brightening (orange patch) over-plotted, (c) Automatically traced $SGPIL_A$ (blue curve) over-plotted on Bz-gradient map, (d) Automatically traced $SSPIL_A$ over-plotted on magnetic shear map, (e) $SGPIL_A$ over-plotted on Bz map, (f) Vector magnetogram with horizontal magnetic field vectors (red, green arrows) over-plotted on Bz map. $SSPIL_A$ is shown with blue curve. It is noted that the flare brightening generally traces SGPIL/SSPIL with differences (at the middle) increasingly in more complex ARs.}
	\label{Fig_PIL_FL_rib}
\end{figure*}
\subsection{Magnetic flux}
The total absolute flux in the AR is computed by $\Phi=\Sigma |B_z|dA$, where $dA$ is the area of observation pixel. The emergence of electric current embedded in magnetic flux plays a main role in triggering the flares \citep{Leka1996, Schrijver2005,Vemareddy2015}. As SGPIL being the proxy for such emerging photospheric electric currents, the flux near to SGPILs, defined by $R$, must have strong relation with the flares as  first proposed by \citet{Schrijver2007}. Following the procedure prescribed in \citet{Schrijver2007}, we constructed the binary maps of positive and negative strong-field masks with the threshold value of +300 G and -300 G respectively \citep{Guerra2018}, then these maps were dilated by a  $6\times6$ square pixel  kernel to create dilated positive and negative bitmaps. Regions of overlap of these two bitmaps were identified as PILs. Then this region of overlap is convolved with normalized Gaussian of FWHM 15Mm to create the weighting map. Finally, this weighting map is multiplied by the absolute value of magnetogram($B_z$) and the weighted unsigned flux density integrated over all pixels gives the value of $R$. $R_{SG}$ is measured similarly along the $SGPIL_M$ segments. Note that $R$ and $R_{SG}$ differ by flare brightening information. One such example is shown in Figure~\ref{Fig_Wmap}.

\subsection{Magnetic energy}
The total energy in the coronal field is estimated using virial theorem equation provided the magnetic field observations at the photospheric surface \citep{Chandrasekhar1961,Molodensky1974,Low1982} and is given by                                      
\begin{equation}
	E = \frac{1}{4\pi} \int_S (x B_x + y B_y) B_z dx dy                
\end{equation}
    
The use of this equation at the solar photosphere is restricted because the photosphere field is not fully force-free and not precisely flux-balanced in a finite area surrounding the AR of interest. Therefore, the photospheric energy estimate serves only as proxy to the energy content in the AR magnetic structure. The potential magnetic fields are derived from the $B_z$ component by using the Fourier transform method \citep{Gary1989}. Since the potential field state is a minimum energy state, by subtracting potential energy ($E_p$) from the total energy ($E$) one obtains the upper limit of the free energy ($E_f$) available in the AR to account for energetic events likes flares and CMEs. 
\begin{figure}
	\centering
	\includegraphics[width=.49\textwidth,clip=]{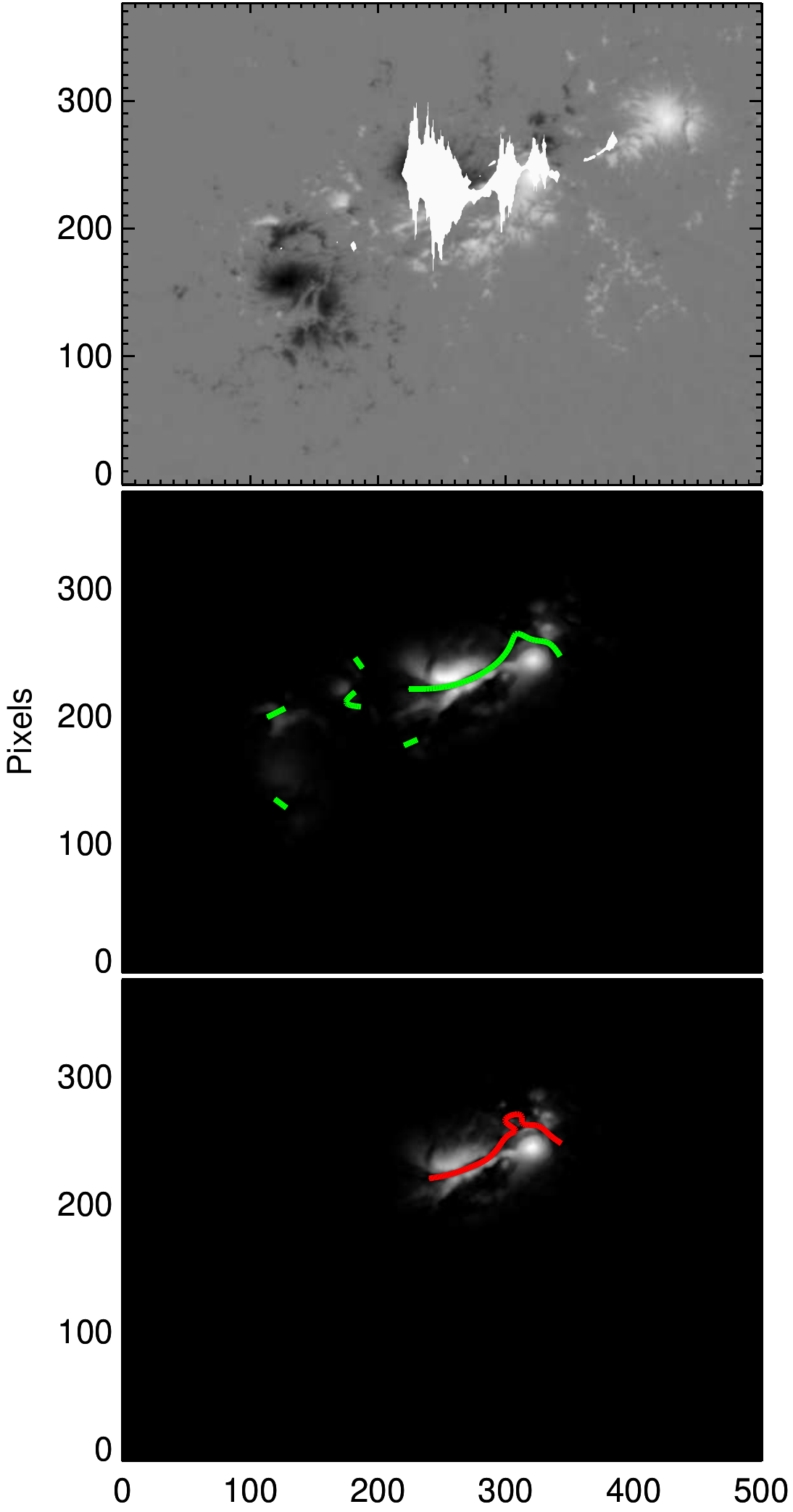}
	\caption{Top: The HMI magnetogram B$_{z}$ map of AR 11158 obtained on 15 Feb 2011 during X2.2 flare. Contours of the AIA 1600~\AA~ flare brightening at flare peak time(01:45 UT) are over-plotted. The unsigned flux density of $B_z$ summed over all pixels gives $\Phi$. Middle: Magnetogram $B_z$ multiplied with the weighted map of the field near $SGPIL_A$(Green curves).  Summing the absolute values of $B_z$ at these pixels yields $R$. Bottom: Magnetogram $B_z$ multiplied with the weighted map of the field near $SGPIL_M$(Red curve). Summing the absolute values of $B_z$ at these pixels gives $R_{SG}$.}
	\label{Fig_Wmap}
\end{figure}

\subsection{Decay index of coronal background field}
The coronal magnetic field gradient is referred to decay index $n(z)$, considered as the important parameter in controlling the eruptiveness of an AR. In the flux rope based CME models \citep{Torok2005,Olmedo2010,Cheng2011}, if the decay index of overlying magnetic field reaches a critical value, then it results in torus instability (rapid decaying of overlying magnetic fields), which leads to CME eruption. 

The decay index is defined as, $n(z) = -\frac{z}{B_h} \frac{\partial B_h}{\partial z}$, where $z$ is the geometrical height from the bottom boundary (photosphere) and $B_h$($=\sqrt{B_x^2+B_y^2}$) is the horizontal field strength. We computed the background field in the entire volume of AR magnetic structure by potential field approximation \citep{Gary1989} using the observed vertical component of the magnetic field at the photosphere. From this extrapolated field, the $B_h$ as a function of height is obtained at eight points along the main PIL and an average of $n(z)$ is then derived. \citet{Torok2005} proposed a constant value of 1.5 as a critical decay index ($n_{crit}$), and is a subjective value based on different conditions. The height at which the $n$-curve reaches the $n_{crit}$ of 1.5 is considered as critical height. We estimated this critical height for all the events in our sample.
\begin{figure*}[!ht]
	\centering
	\includegraphics[width=.49\textwidth,clip=]{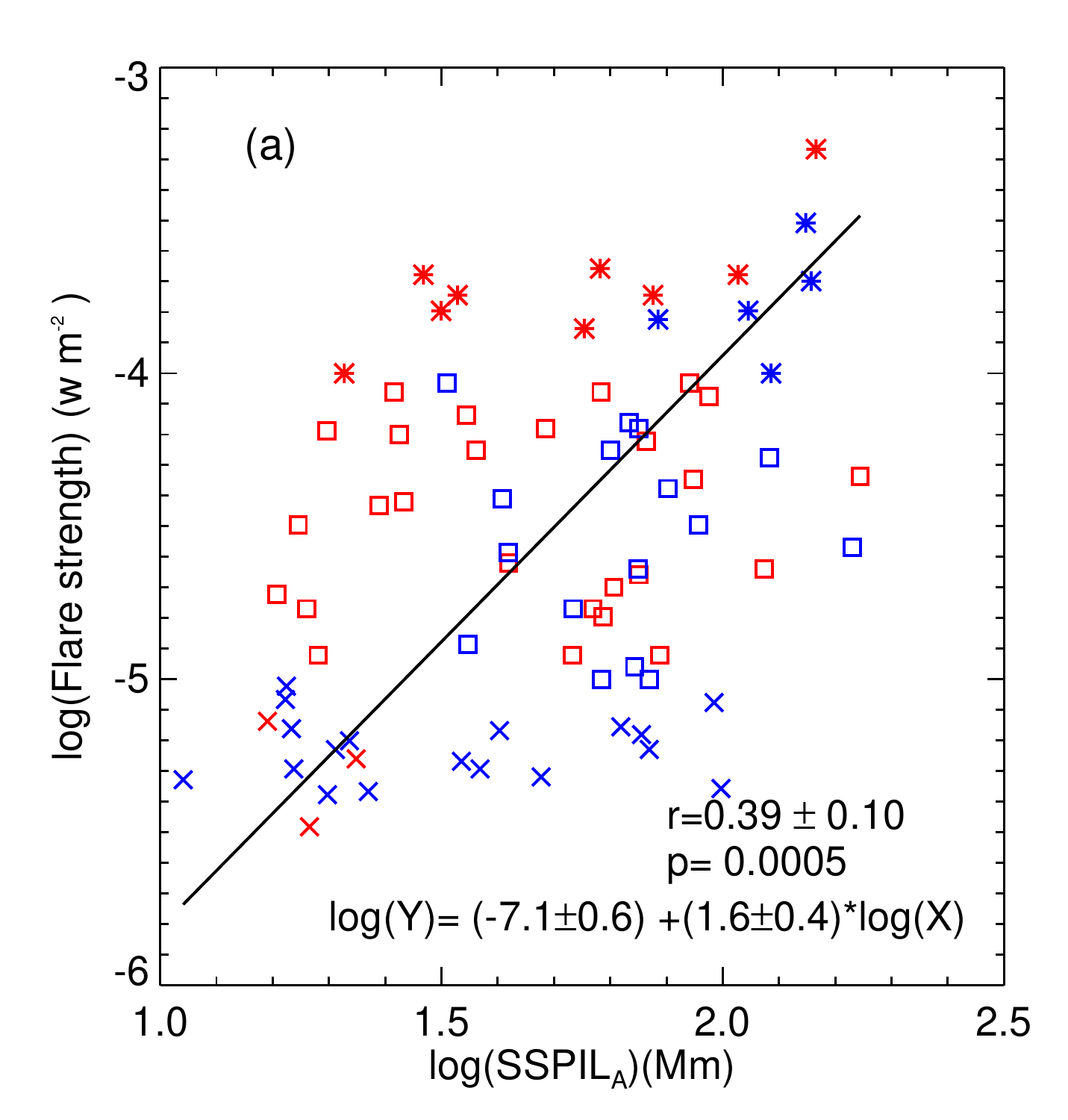}
	\includegraphics[width=.49\textwidth,clip=]{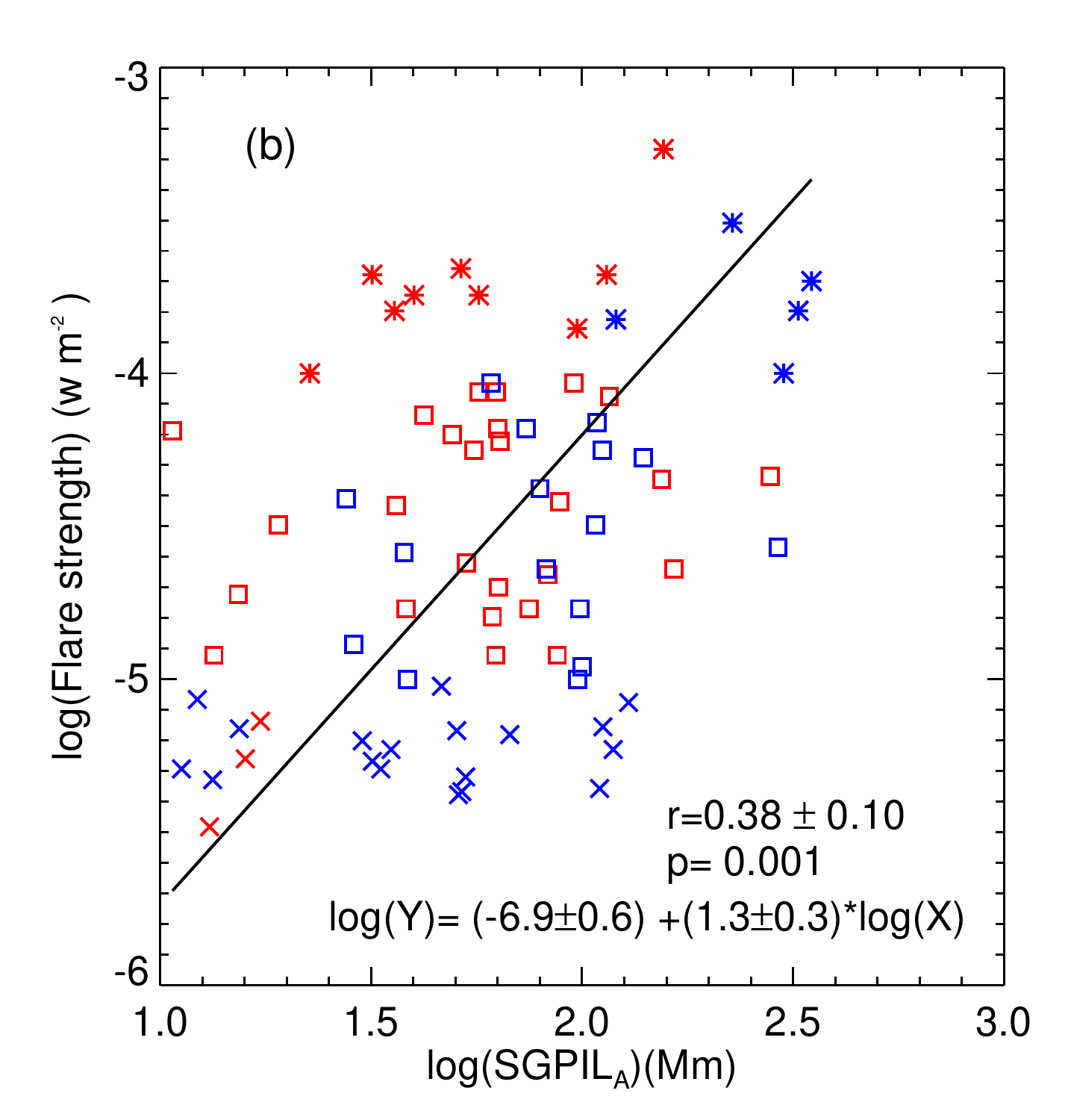}
	\includegraphics[width=.49\textwidth,clip=]{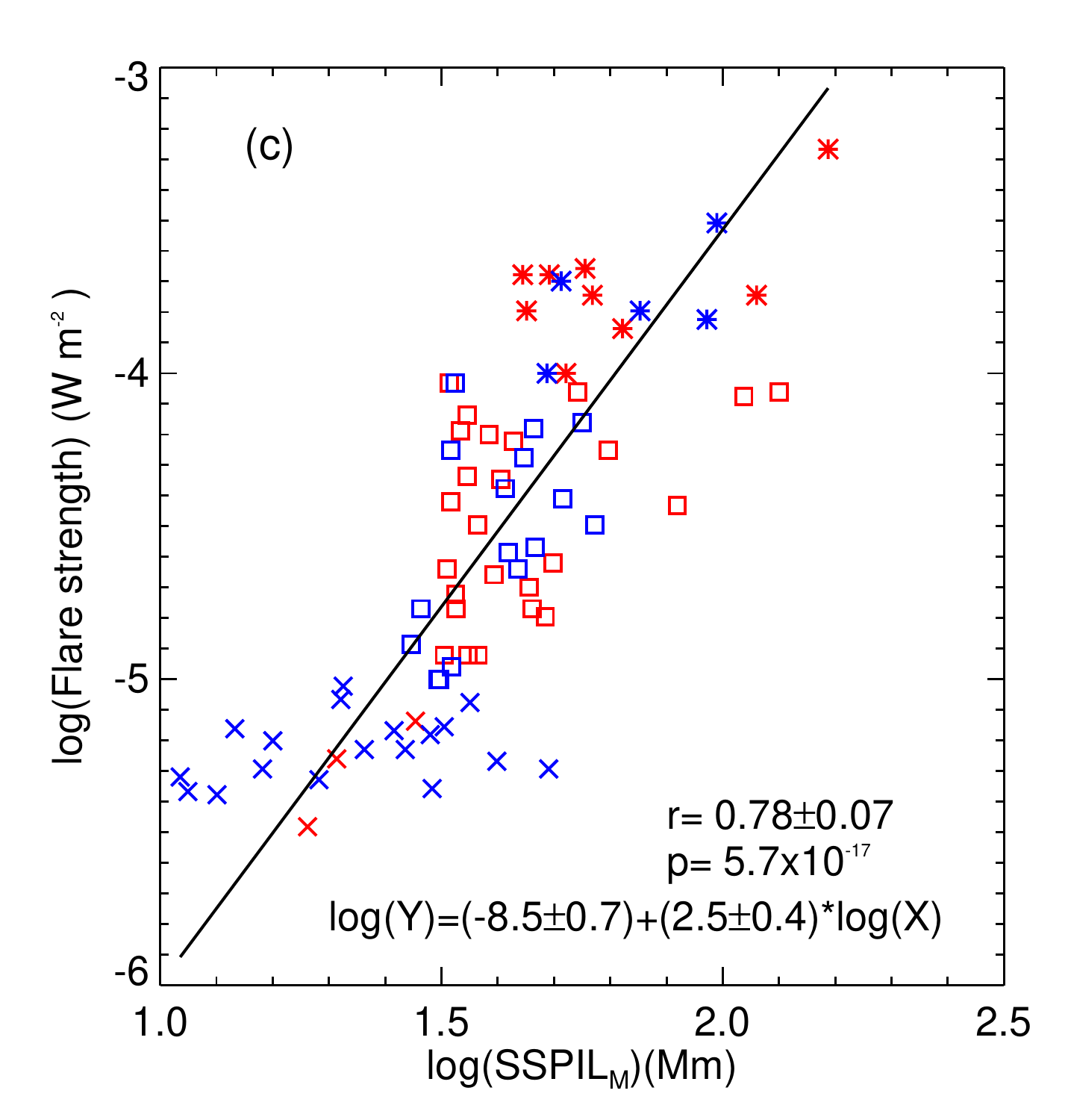}
	\includegraphics[width=.49\textwidth,clip=]{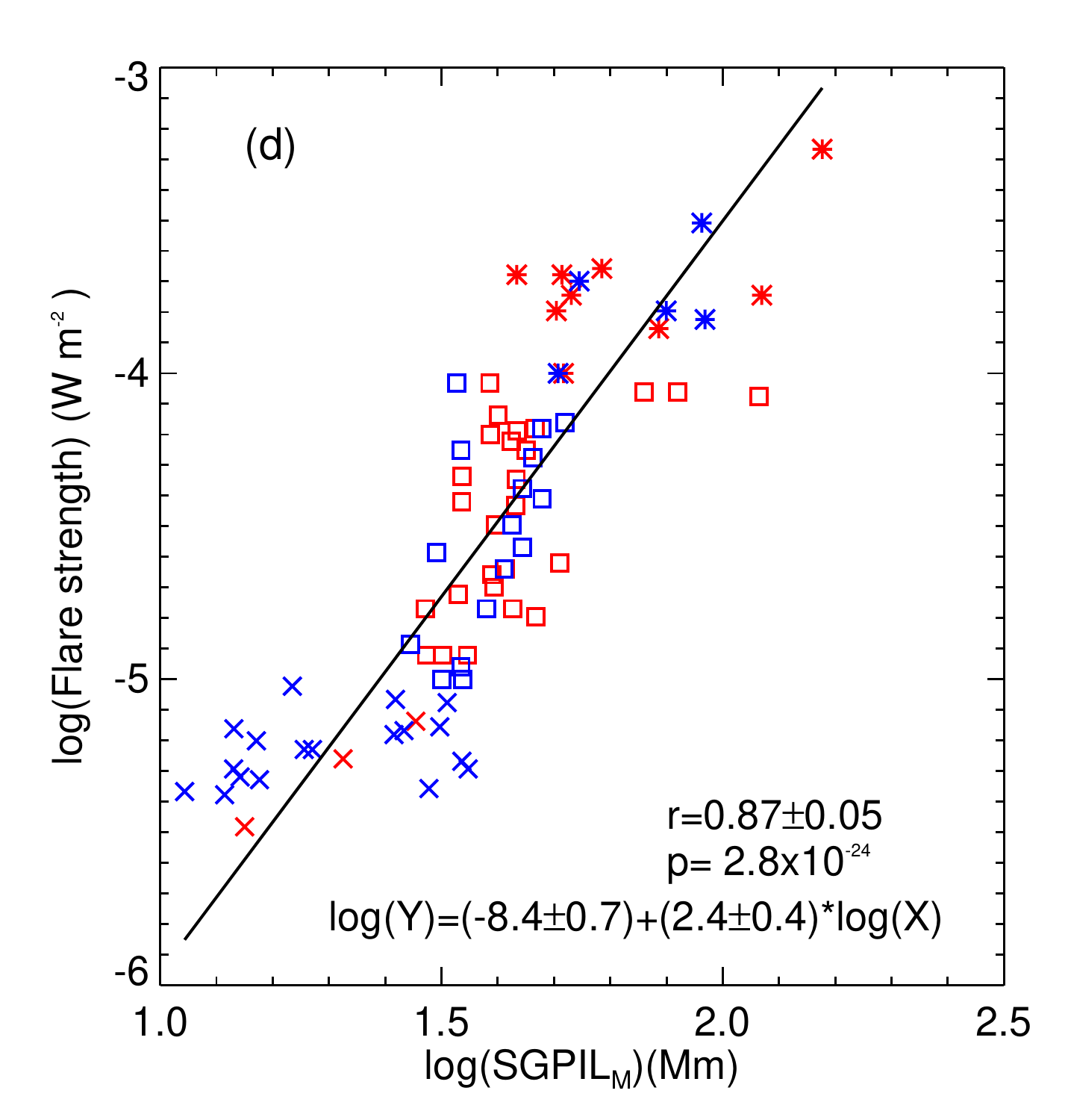}
	\caption{Top two panels display the scatter plots of automatically detected $SSPIL_A$ and $SGPIL_A$ with flare strength and bottom two panels represents the scatter plots of manually detected $SSPIL_M$ and $SGPIL_M$ respectively. Spearman ranking correlation coefficient ($r$), two sided significance(p-value) and the equation of solid fitted line are inserted in respective panels. The cross, square and asterisk symbols represent the C-class, M-class and X-class flares respectively. Red (blue) color of these symbols correspond to eruptive (confined) flare cases. {Note that manual detection method implies a higher correlation to flare strength.}}
	\label{Fig_PIL_len}
\end{figure*}

\section{Results}
\label{res}
\subsection{PIL Length Versus Flare Strength}
High vertical field gradient and strong shear usually appear in the vicinities of PILs, where flares frequently occur \citep{Hagyard1986,Hagyard1988,Falconer2003,Sharykin2016}. 
Following the procedure described in Section 2.1, we estimated the lengths of both SSPIL and SGPIL for all the sample flaring ARs by automatically called as $SSPIL_A$ \& $SGPIL_A$ and manually as $SSPIL_M$ \& $SGPIL_M$ respectively. In this regard, we used vector magnetograms available immediately before the initiation time of the flares. In Figures~\ref{Fig_PIL_len}(a \& b), we showed the relationship of $SSPIL_A$ and $SGPIL_A$  with the flare strength. As our algorithm identifies multiple PIL segments of high gradient and strong shear in an AR and their summation $SGPIL_A$ and $SSPIL_A$ respectively shows weak correlation with the flare strength at a {Spearman's rank correlation coefficient $\sim$0.4. As we are keen to find the correlation between the flare strength and  PIL segment of high gradient and strong shear in the flare brightening region, we traced $SGPIL_M$ and $SSPIL_M$ manually as described in the Section 2.1. 
\begin{figure*}[!ht]
	\centering
	\includegraphics[width=.99\textwidth,clip=]{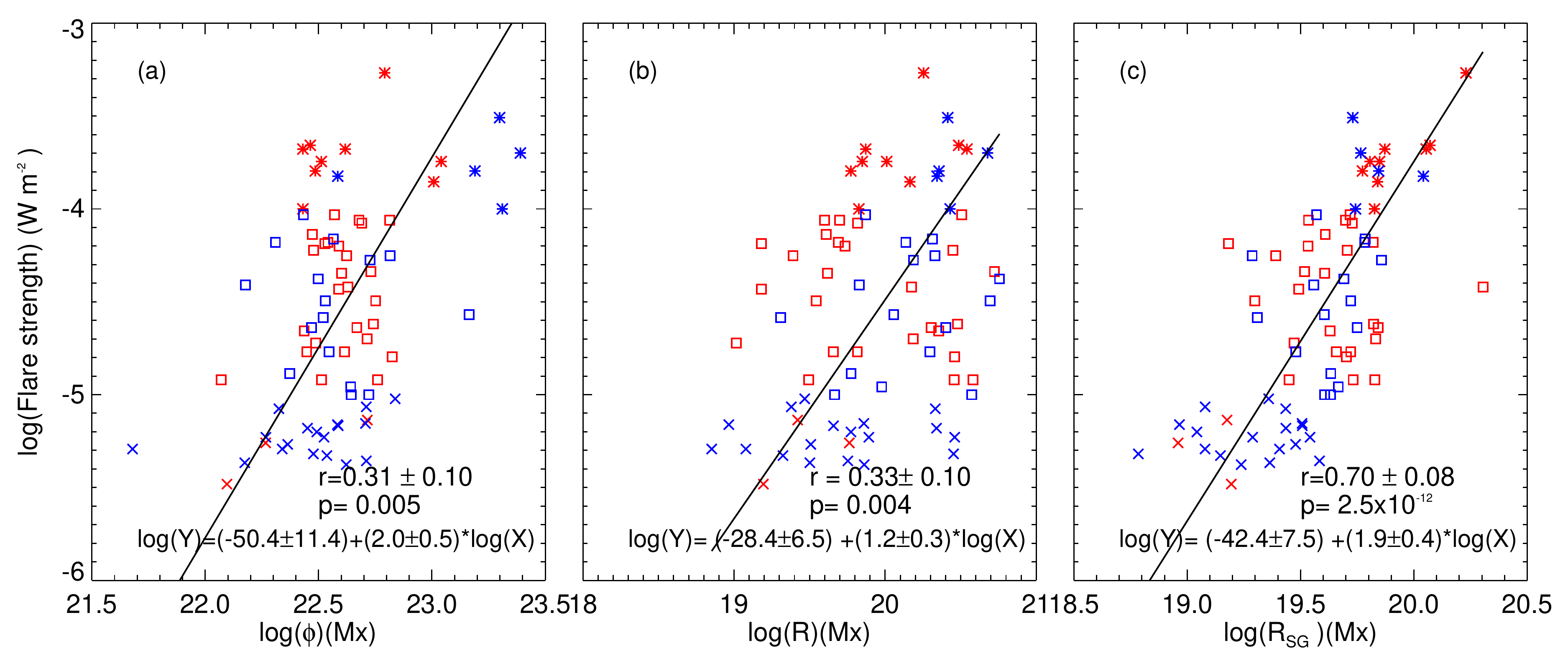}
	\caption{Scatter plots of $\Phi$, $R$ and $R_{SG}$ with flare strength in left, middle, and right panels, respectively. Spearman ranking correlation coefficient, two-sided significance(p-value) and the equation of solid fitted line are inserted in respective plots. The cross, square and asterisk symbols represent the C-class, M-class and X-class flares respectively. Red (blue) color of these symbols corresponds to eruptive (confined) flare cases.}
	\label{Fig_flux}
\end{figure*}

Here and in the following studies, linear regression analysis is done by using \texttt{FITEXY.PRO} routine available in SolarSoftWare library, where it uses least-square approximation in one-dimension to fit the best straight line to the data with errors in both coordinates. We used standard deviation of both coordinates as error inputs. The uncertainties in both the obtained coefficients are shown in the equation, which are inserted in their respective figure panels. We estimated the Spearman ranking correlation coefficients ($r$) in all our studies and the  standard error in r is estimated by  $ERR_r = \sqrt((1-r^2)/(n-2))$. From here onwards, we refer Spearman ranking  correlation coefficient as correlation coefficient (CC).

\begin{figure*}[!ht]
	\centering
	\includegraphics[width=.49\textwidth,clip=]{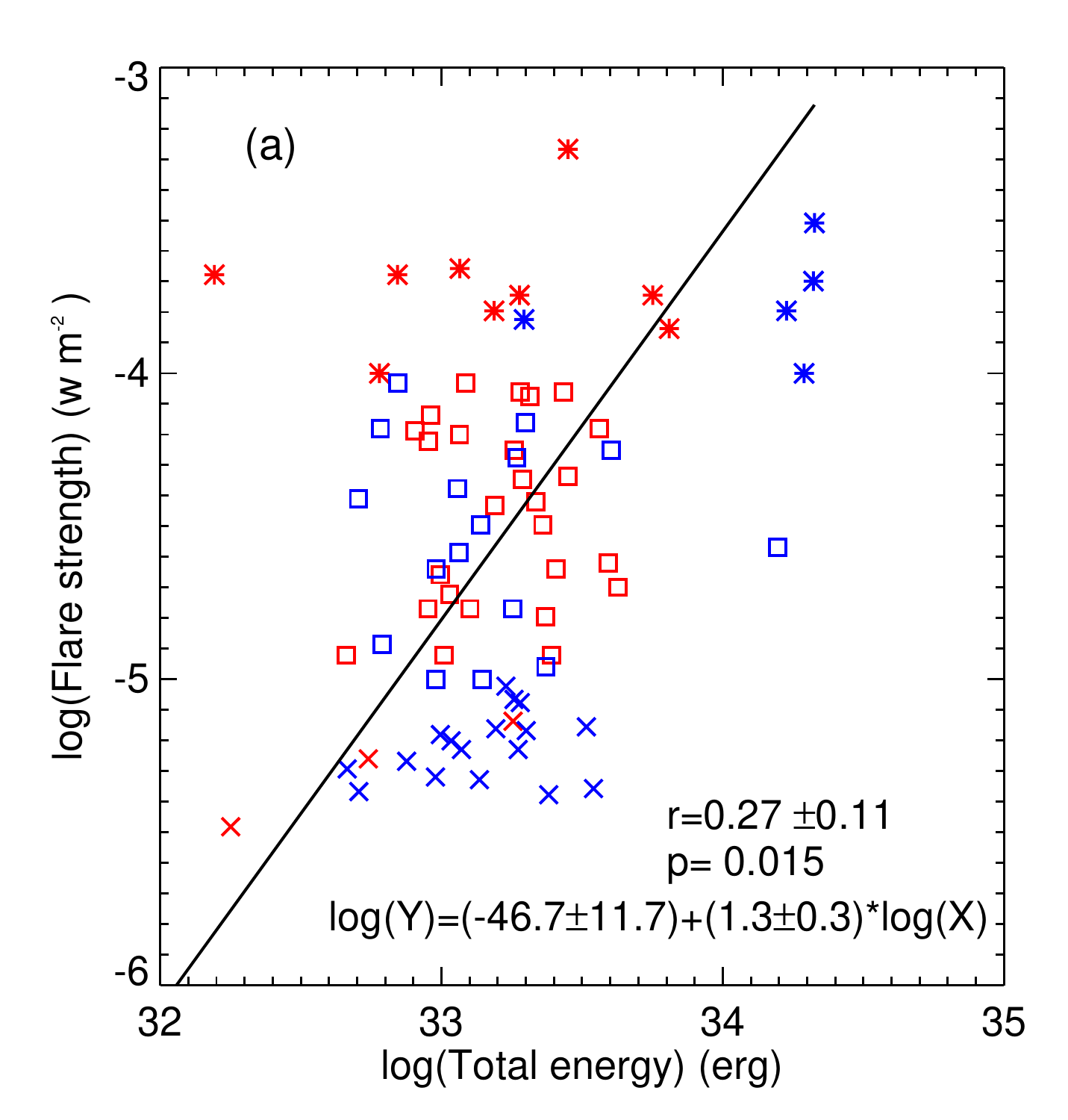}
	\includegraphics[width=.49\textwidth,clip=]{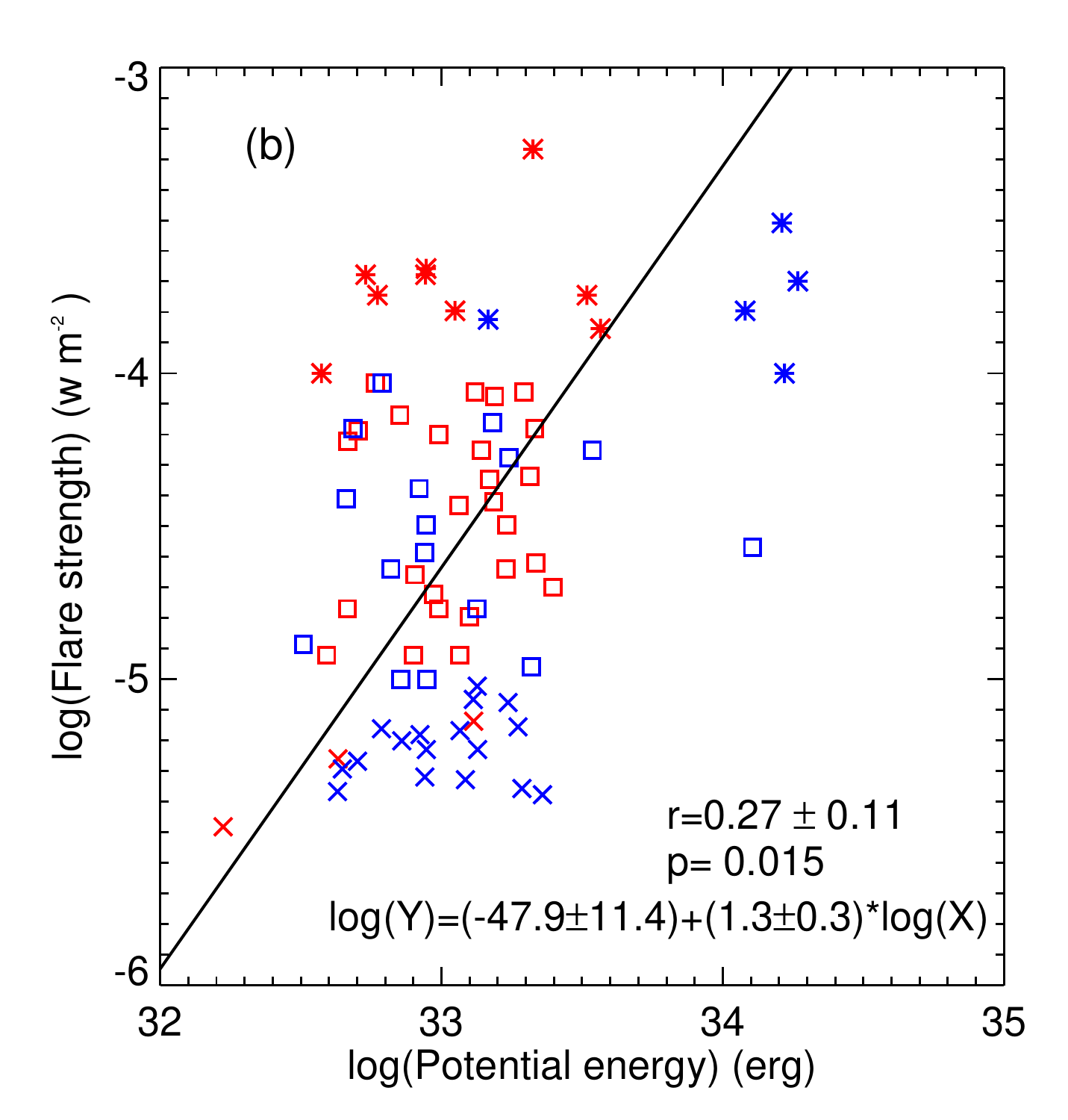}
	\includegraphics[width=.49\textwidth,clip=]{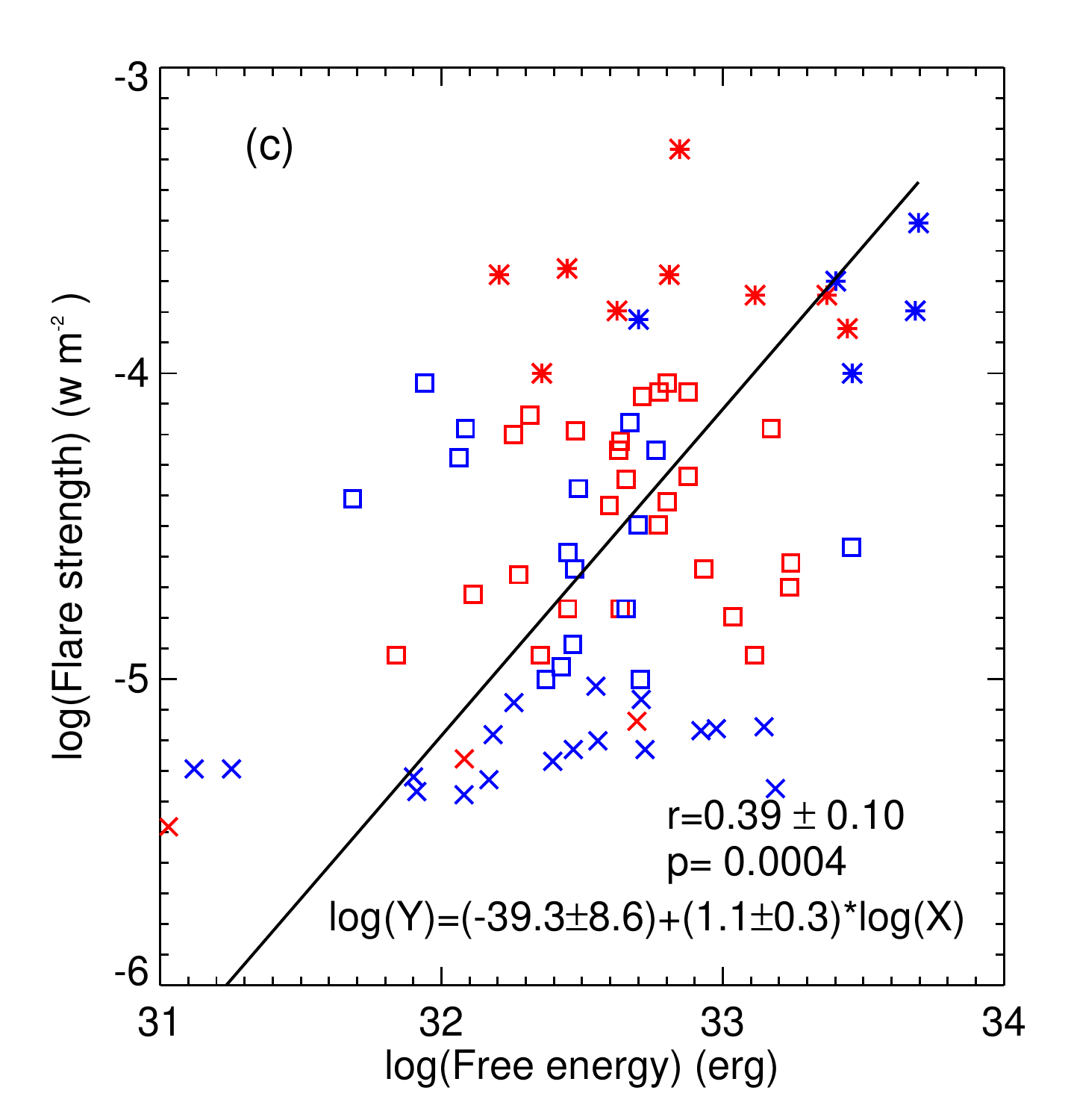}
	\includegraphics[width=.49\textwidth,clip=]{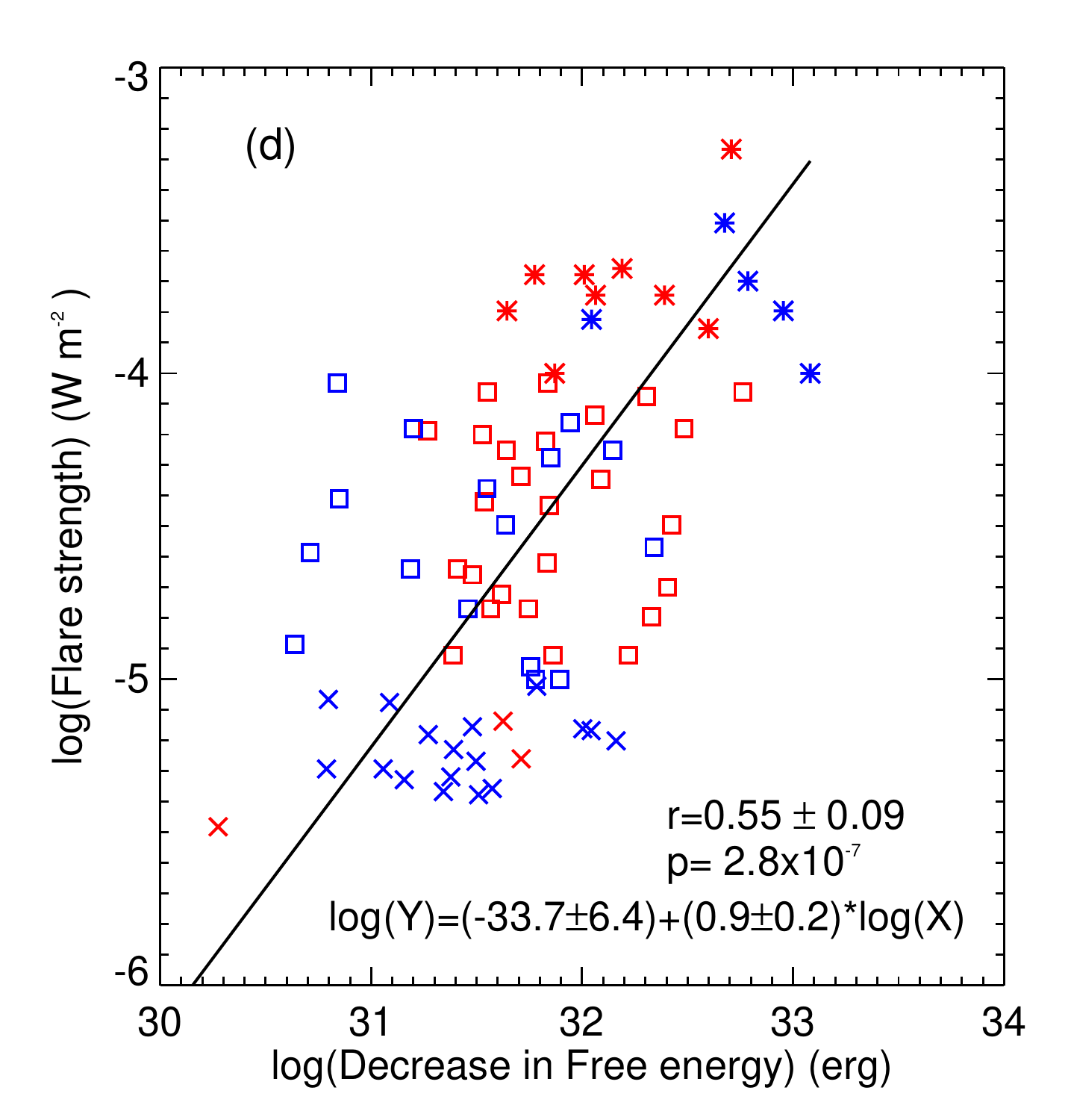}	
	\caption{Scatter plots of a) total magnetic energy, b) potential energy, c) free energy, d) decrease in magnetic free energy, against the flare strength in logarithmic scale. Spearman ranking correlation coefficient, two sided significance(p-value) and the equation of solid fitted line are inserted in respective panels. The cross, square and asterisk symbols represent the C-class, M-class and X-class flares respectively. Red (blue) colors of these symbols corresponds to eruptive (confined) flare cases. Note a higher correlation of decrease in free energy with flare strength.}
	\label{Fig_ene}
\end{figure*}

In Figures~\ref{Fig_PIL_len}(c \& d), the $SGPIL_M$ and $SSPIL_M$ are plotted against the flare strength, respectively. It can be seen that the length of both $SGPIL_M$ and $SSPIL_M$  increase with flare strength. It indicates that more intense flares tend to occur from ARs having larger PILs weighted by strong vertical field gradients, a proxy for strong shear. ARs with large PILs are indicative of complex field structure, and PILs with large gradients are indicative of shearing. Therefore, the SGPIL and SSPIL both describe the global non-potentiality of ARs. $SGPIL_M$ has relatively better correlation with the flare strength at a CC of 0.87 than that of $SSPIL_M$ with flare strength at a CC of 0.78. These strong correlations with flare strength confirms that the flare productivity depends on the non-potentiality of ARs. Importantly, it is noted that the 35 out of 38 CME associated cases (red symbols) have SGPIL length larger than a threshold of $Log10(SGPIL)=1.5$ (31Mm). This may indicates that the CME occurrence requires a certain minimum SGPIL length for being eruptive from confined environment.

\begin{table*}
	\centering
	\caption{Contingency table based on $SGPIL_M$ length for the forecast of CMEs}
	\begin{tabular}{|c|c|c|c|}
		\hline 
		No. of samples	& CME occurrences & Non-CME occurrences &  Total \\ 
		\hline 
		Log($SGPIL_M$) $\ge$ 1.5 (yes)	& 35 (H) & 22 (FA)  &  57		\\ 
		Log($SGPIL_M$) $<$ 1.5 (No)	& 3 (M) & 17 (CN)  &  20       \\ 
		\hline
		Total	&  38 & 39  &  77								\\ 
		\hline 
	\end{tabular} 
	\label{tab1}
\end{table*}

The above relation is tested for statistical significance. Contingency table for $Log(SGPIL_M) \ge  1.5$ and CME productivity from all the flaring ARs (both confined and eruptive) in our sample is constructed and shown in Table 1.
Fisher's exact test is applied to check for the significance of relationship between these two variables. P-value is determined by using hyper geometric distribution and its value is found to be 2.9$\times$10$^{-4}$, which is statistically significant. This relation is not significant for $SGPIL_A$, as we can see there is no clear threshold value for distinguishing the confined and eruptive flares. Further, this threshold value of $SGPIL_M$ is used as forecasting parameter for CME productivity. Forecast verification measures like True Skill Statistic (TSS) and Heidke Skill Score(HSS) were calculated using the forecast contingency table \citep{Bloomfield2012} as shown in Table~\ref{tab1}. Both of these measures account for the correct random forecasts and can have the range of skill scores from -1 to +1. TSS and HSS were found to be 0.357 and 0.355 respectively.

Further, the estimated length of the $SSPIL_M$ is plotted against that of $SGPIL_M$ (plot is not shown). They are highly correlated at a correlation coefficient of 0.92. Our result in a large sample is consistent with \citet{WangHM2006}, who showed a strong correlation between magnetic gradient and magnetic shear in primary PILs of five strong flaring ARs.  This high correlation is not a surprise as \citet{Falconer2003} already indicated that SGPIL is a viable proxy for SSPIL and it is well correlated with the CME productivity of ARs. As SGPIL can be measured from line-of-sight magnetogram, it serves as a substitute for SSPIL in CME/flare forecasting studies. 

\subsection{Total unsigned flux, Magnetic energy versus Flare strength}

In the past many studies conducted on the association of flares and the flux content of ARs \citep{Leka2003a,Leka2003b,Schrijver2007,Bobra2015,Kazachenko2017}. They all commonly found a weak correlation between  total unsigned flux of whole AR with the flare strength. Recently, \citet{Kazachenko2017} found strong correlation between reconnection flux i.e., the total unsigned flux of flare ribbon area with the flare strength. In the same line, we conducted three different studies to find the relationship between flares and the total unsigned flux quantitatively.  Firstly, the total unsigned flux ($\Phi$) of whole AR, secondly Schrijver's $R$ value which is the total unsigned flux within 15Mm of all SGPILs in the AR and thirdly the $R_{SG}$ (similar to the flare ribbon reconnection flux as defined in \citealt{Kazachenko2017}), which is the total unsigned flux within 15Mm of $SGPIL_M$ but the only the flare ribbon extent. 
\begin{figure*}
	\centering
	\includegraphics[width=.98\textwidth]{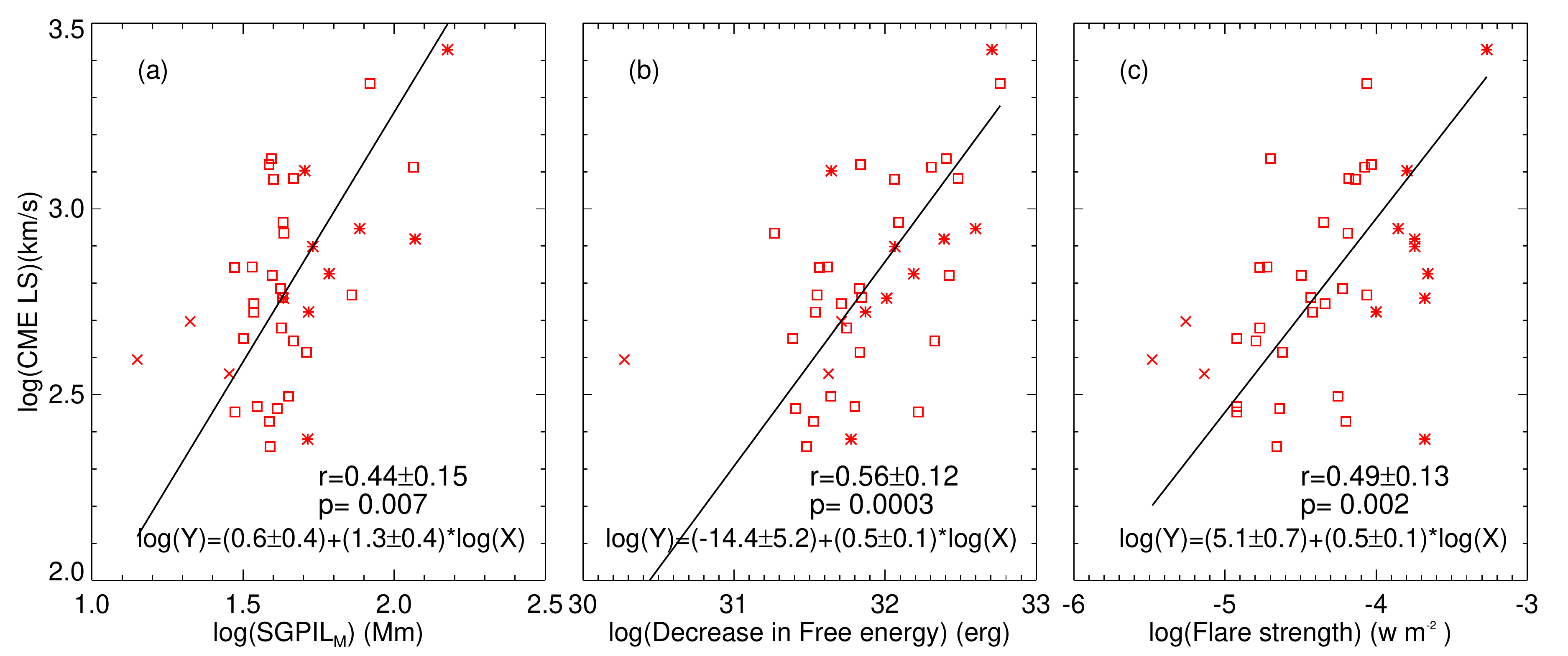}
	\caption{Relation of CME kinematics on AR magnetic properties. The scatter plots of a) $SGPIL_M$, b) decrease in amount of free energy and c) flare strength with CME speed. Spearman ranking correlation coefficient, two-sided significance(p-value) and the equation of solid fitted line are inserted in respective plots. }
	\label{Fig_CMEkin}
\end{figure*}
In Figure~\ref{Fig_flux}a, the total unsigned flux from the whole AR is plotted against the flare strength and shows a weak correlation of CC$\sim$0.31 . In Figure~\ref{Fig_flux}(b) , $R$ value is plotted against the flare strength which also shows a weak positive correlation of CC=$\sim$0.33. Whereas in Figure \ref{Fig_flux}c, $R_{SG}$, the flux from  the flare eruption site, has strong association with flare strength of CC $\sim$0.70. These results indicate that the flux near to $SGPIL_M$ has strong connection to the flare strength. This suggests that flare strength neither depends on the total unsigned flux of whole AR as it just represents the size of the AR nor depends on the flux near to all strong gradient PILs in an AR but there is a strong physical dependence with the flux from the region of 15Mm about $SGPIL_M$ which overlaps the flare brightening. This result also confirms \citet{Kazachenko2017} result and may be more robust given better statistics. This refined relation points the difficulty of predicting the flare and its strength based on the pre-flare magnetograms. 

Further, we explored the relationship between magnetic energy and flare strength in Figure~\ref{Fig_ene}. The total magnetic energy for all flare events in our sample at initial and end timings of flares were estimated (Section 2.3). The total magnetic energy and potential energy at the initial times are weakly correlated with the flare strength of correlation coefficient of 0.27 each as shown in panels~\ref{Fig_ene}(a \& b).  In panel~\ref{Fig_ene}(c), the magnetic free energy is plotted against the flare strength. Despite the points are scattered, there is a moderate positive correlation with a CC of 0.39. 

In all our sample, the magnetic free energy estimated at flare end timings are found to be smaller than the magnetic free energy estimated at the flare initial timings. Panel~\ref{Fig_ene}(d) depicts the correlation between the decrease in free energy with the flare strength. Unlike the energy estimates at initial times, the scatter plot shows a moderate positive correlation ($CC=0.55$) indicating a physical link between the decrease in amount of free magnetic energy and intensity of flares. It indicates that the total amount of magnetic energy and potential energy possessed by an AR has very little to do with the intensity of flares that the AR has produced but the difference in total magnetic free energy before/after the flare is directly proportional to the strength of the flare, as one would expect. This result is corroborated by the strong correlation between flare strength and amount of free energy decreased during the flares \citep{Vemareddy2012}, and hence it is clearly evident that larger the amount of energy released, stronger the erupted flares. Notedly, the free energy decrease for the CME cases (35 of 38, red symbols) starts from  $2\times10^{31}$ ergs ($Log(31.3)$) and may indicates the threshold energy required for a flare to be eruptive. 
\begin{figure*}
	\centering
	\includegraphics[width=.97\textwidth]{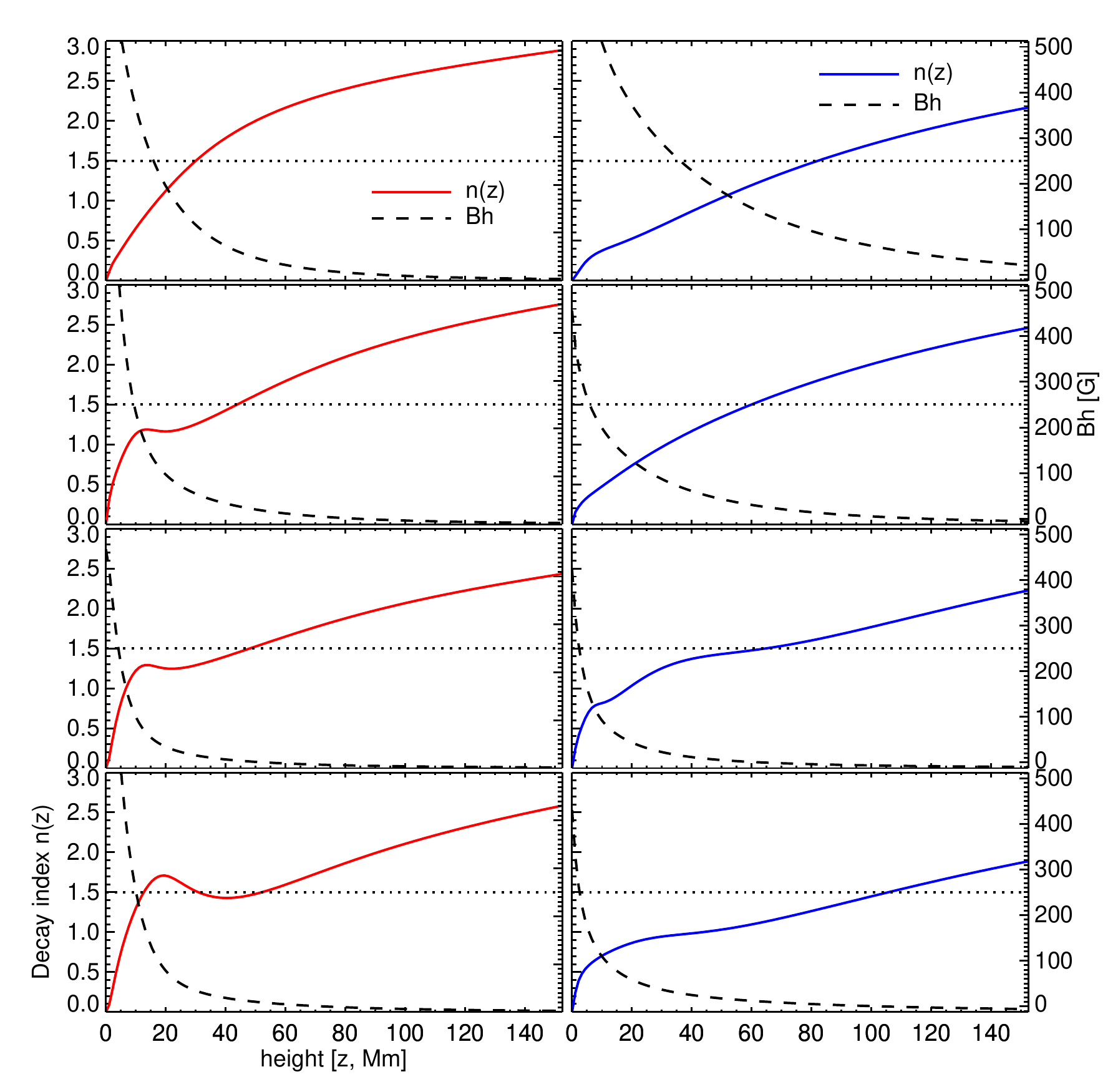}
	\caption{Plots of decay index of background horizontal field strength as a function of height. \textit{Left column:} Typical cases of eruptive flares M4.5, X2.2, X1.0 and X2.1 originated from ARs 12158, 11158, 12017 and 12297 respectively. The critical heights in these events are noted as 30.21Mm, 42.4Mm, 40.94Mm and 10.96Mm respectively.  \textit{Right column:} Typical cases of the confined flares X2.0, M1.1, C5.4 and C6.9 originated from ARs 12192, 12253, 11936 and 12472 respectively. The observed critical heights for these cases are 82.25Mm, 57.76Mm, 62.51Mm and 103.1Mm respectively. $B_h$ is also shown with the y-axis scale on the right in each panel. The dotted horizontal line refers to the $n_{crit}=1.5$.}
	\label{Fig_dc}
\end{figure*}
\subsection{Magnetic properties of Active region versus CME kinematics}

From the sample of 77, 38 flares are CME associated (eruptive flares) and have known source ARs. We used the online LASCO/CME catalog to determine the CME/Flare association and linear speed (LS) of the CMEs. The relationship between these 38 source AR properties and the associated CME speeds are examined in Figure~\ref{Fig_CMEkin}. The observed CME speed follows the positive correlation with $SSPIL_M$ and $SGPIL_M$ with a correlation coefficients of 0.43 and 0.44 respectively. For reference, only the plot of $SGPIL_M$ versus CME linear speed is displayed. This implies that faster CMEs tend to initiate in ARs with longer lengths of PIL surrounded by greater magnetic complexity \citep{Falconer2003,Song2006}.

There is reasonably high positive correlation between the decrease in magnetic free energy after the flare eruption and CME linear speed (panel~\ref{Fig_CMEkin}(b)) with a CC of 0.56.  We explored the relationship between magnetic free energy(FE) before the flare eruption and CME linear speed as well but the correlation is weak (plot is not shown). As the decrease in free energy involves the difference ($FE_{initial}-FE_{end}$), it correlates well with CME speed. Further, we also studied the relationship between total unsigned flux with CME linear speed but found out weak correlation between them (plot is not shown). These results suggest that there is a strong physical link between the released non-potential energy of source AR and CME speed but there is no evidence that CME kinematics depends on the size of the ARs. This result is in accordance with the past studies that were done using line-of-sight magnetograms \citep{chen2011JGRA}. \citet{chen2011JGRA} claimed that the size, strength and complexity of ARs do little with the kinematic properties of CMEs, but have significant effects on the CME productivity.

Further, flare strength and CME speed are also positively correlated with a CC of 0.49, shown in panel~\ref{Fig_CMEkin}(c). This suggests a general relation that CME speeds are proportional to flare strength and would underly the action of impulsive reconnection on the expelled CME in agreement with previous studies \citep{Guo2007,YumingWang2007} but with few recently found exceptions (like \citealt{Sun2015}).

\subsection{Confined and Eruptive Flares}
In this section, we investigated the role of background coronal field in the confined flares (without CMEs) and eruptive flares (associated with CMEs). Following from procedure described in Section 2.4, we estimated the critical decay index heights for all 77 events in our sample. In Figure~\ref{Fig_dc}, $n(z)$ is plotted for typical cases of eruptive (left column panels) and confined flares (right column panels). $B_h$ (average of 8 points along main PIL) as a function of height is also shown in the respective panels. For eruptive panels, the extrapolated field reached this critical value at a height below 45Mm. Notedly, at about 10Mm the curve exhibits a gradual steepness with a "bump" in the eruptive cases and is missing in confined cases with almost a smooth curve. The "bump" is also found in \citet{Cheng2011} study and is interpreted as a distinct shape for eruptive and non-eruptive flare events. It was noticed that the $B_h$ decreases faster in the low corona (about 10Mm), and the appearance of a "bump" in the eruptive flare cases which may indicate the ability of the twisted flux (flux rope) to experience torus instability. From these panels, it is seen that the $n(z)$ curve is steeper reaching $n_{crit}=1.5$ within 40Mm for eruptive cases. For confined ones, the $n(z)$ is gradually reaching $n_{crit}$ well beyond 50Mm height. 

In Figure~\ref{Fig_dc_ch}, we plot the critical heights of all events versus flare strength. It can be seen that the both the confined (blue circles) and eruptive (red circles) spread all along the $n_{crit}$ height and have no relation to the flare strength. Further, \citep{Liu2008} proposed the typical height of eruption onset is 42$\pm$1 Mm, based on the average initial heights of the 4 observed filament events consisting of two failed eruptions and two full eruptions. Meanwhile, many recent studies of full eruptions indicate a critical height well below 42Mm (eg., \citealt{Cheng2011,Vemareddy2014,Sun2015}). Based on this segregation of the events (vertical dashed line), about 90\% (34 of 38) of the eruptive flares have the critical height less than 42 Mm and nearly 70\% (27 of 39) of the confined flares have the critical height beyond 42Mm as evident in Figure~\ref{Fig_dc_ch}. Though the dependency of critical heights is a matter of individual cases, generally critical heights depend on the strength of background field confinement but does not depends on the intensity of flares. This indicates that the background field for confined cases has extended or stronger confinement than the eruptive ones. Depending on the extent of this confined environment, the unstable core field (or flux rope) near the main PIL would be suppressed or become a CME and is the subject of individual cases. From this statistical study, we propose that a CME is likely from an AR coronal background field where $n(z)$ reaches $n_{crit}$ below 42Mm.  
\begin{figure}
	\centering
	\includegraphics[width=.49\textwidth]{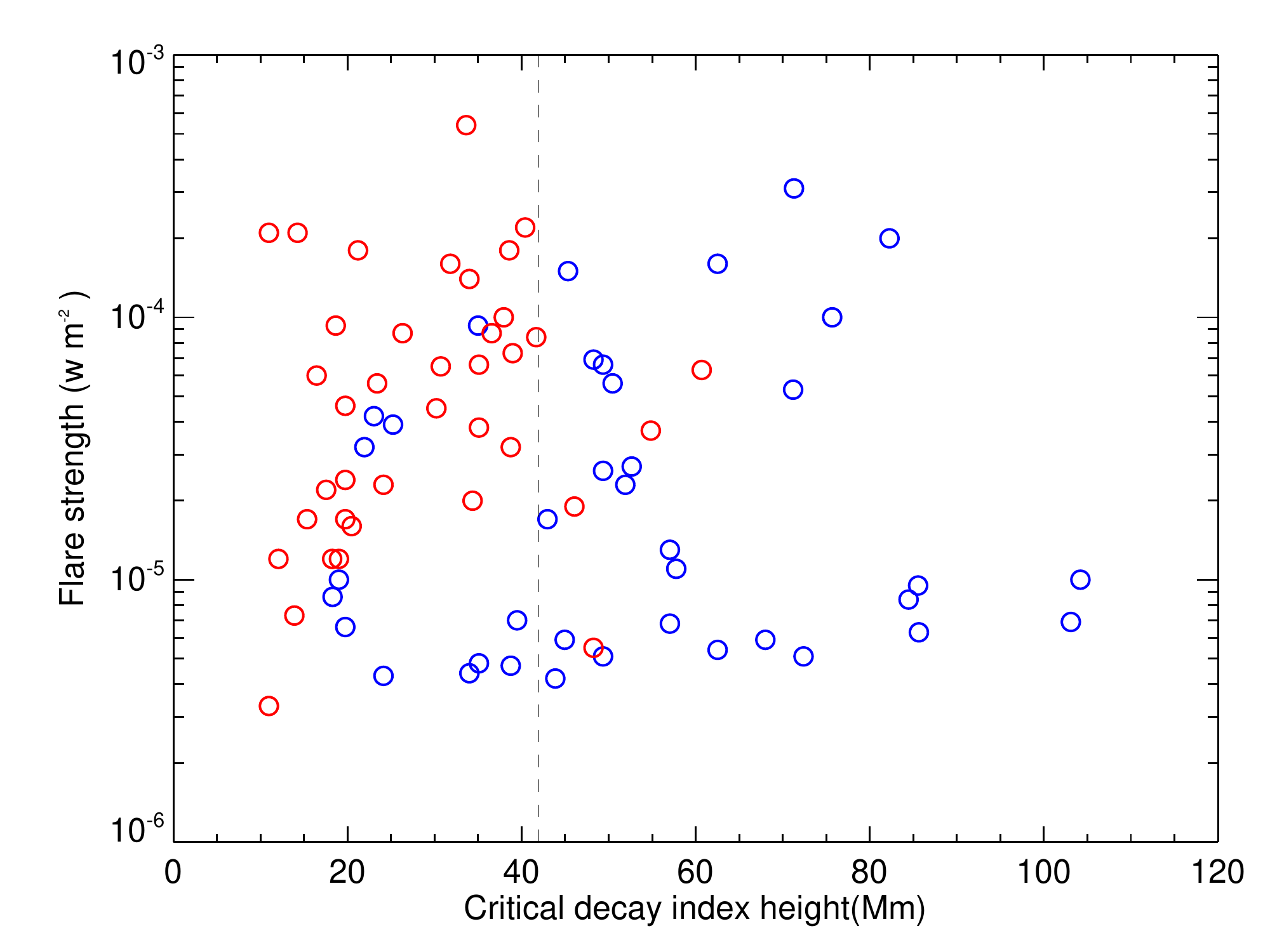}
	\caption{Scatter plot of critical heights for all the flare events in the sample with the flare strength. Red (blue) circles represent the eruptive (confined) flares. Vertical dashed line refers to critical height of 42Mm and divides the eruptive and confined flares. \label{Fig_dc_ch}}
\end{figure}
\section{Summary and Discussion}
\label{summ}
Using the HMI vector magnetic field observations, we have studied the relation of degree of magnetic non-potentiality with the observed flare/CME in ARs. Studying the relation of these properties and establishing statistically significant link with the observed activity is the key in the flare/CME forecasting models. In this connection, we made a systematic analysis of several non-potential proxies, including decrease in free magnetic energy proxy, during flares/CMEs of different magnitude. The chosen flare cases originated from 40 ARs, of which 83\% (19 of 23) in the southern hemisphere, 70\% (12 of 17) in the northern hemisphere follow the dominant helicity sign rule \citep{Pevtsov1995,bao2002}.

The automatically detected $SGPIL_A$ and $SSPIL_A$ lengths have weaker positive correlation with the flare strength (CC=0.40) than the manually detected ones ($SGPIL_M$ and $SSPIL_M$). Manual detection accounts the AIA 1600\AA~flare ribbon extension along PIL. Further, the $SGPIL_M$ have stronger positive correlation with flare strength (0.9) than that of $SSPIL_M$ with flare strength (CC=0.8) and therefore the magnetic gradient seems to be better correlated with intensity of solar flares than magnetic shear. It is in a quantitative agreement with \citet{WangHM2006} where they found this relation in a sample of 5 X-class flaring ARs. 

The total unsigned flux of the entire AR ($\Phi$) and Schrijver's R value were found to be weakly correlated with the flare strength but $R_{SG}$, the total unsigned flux within 15 Mm of $SGPIL_M$ has a statistically significant correlation with flare strength (0.70). This strong correlation signifies the physical link between the PIL flux and flare intensity. The flux near $SGPIL_M$ must be contributing in flaring process similar to the flare ribbon reconnection flux defined by \citet{Kazachenko2017}. As the amount of flux involved in the flaring process increases, the intensity of flares also increases but this effect is camouflaged in $\Phi$ and $R$ value in order to reflect in the intensity of the flares. Both the $SGPIL_M$ and $R_{SG}$ are related to the actual flux involved in the reconnection along PIL. Therefore, it is difficult to predict the flare strength a priori based on pre-flare magnetogram and points the missing key aspect in the flare prediction models \citep{Mason2010}. 

Total magnetic energy and potential energy of flaring ARs derived from virial theorem were found to be weakly correlated with flare strength whereas the magnetic free energy derived at the initial time of flare events has positive correlation with flare strength. Importantly, magnetic free energy decreases after the flare eruption and the amount of decrease in free energy has the strong positive correlation with flare strength. These results suggest that there is a strong physical link between released magnetic free energy and flare productivity, and also the intensity of flares produced. The amount of total magnetic energy and potential energy possessed in an AR are not much related to the intensity of flares that the AR has produced but how much free energy released, has a major contribution to the intensity of flares.

Moreover, we analysed the dependence of CME kinematics from our flaring sample that are associated with CMEs. Both $SSPIL_M$ and $SGPIL_M$ are moderately correlated with CME speed. Also, the amount of magnetic free energy decreased during flare eruptions has relatively strong correlation with CME speed at correlation coefficient of 0.56. These findings imply a general relation that stronger the measures of non-potentiality of source ARs, larger the CME speed \citep{YumingWang2007}. And we also found the most common relation between flare strength and CME speed (CC=0.49), indicating that faster CMEs tend to be associated with more intense flares. 

In addition, background field appears to be key factor for a flare to be eruptive. In 90\% of eruptive flares, the $n(z)$ curve is steeper reaching $n_{crit}$ within 42Mm, whereas $>$70\% confined flares occur in ARs of $n_{crit}$ beyond 42Mm. Recent study by \citet{Vemareddy2017b} inferred the successive sigmoid formation and eruption in AR 12371 under the slow evolving conditions of predominant negative helicity flux injection from the SGPIL region. Minimum length of the SGPIL may be the signature of the twisted flux rope. In the confined AR 12192 (no CMEs but with X-flares), the magnetic flux normalized helicity flux is smaller by a factor of 10 and has no signatures of twisted flux rope in coronal imaging observations. As the flux rope is a continuous bundle of twisted field structure, its existence also assumes continuous SGPIL, but not as small distributed segments. Therefore, besides with the minimum required SGPIL length (31Mm, Figure~\ref{Fig_PIL_len}), the weak background field ($<42$Mm) are suggested to be the prime factors for a flare to be eruptive. 

The above inference is tested for the statistical significance. The skill scores estimated in section 3.1 suggests that $SGPIL_M$ does not have the ability to aptly predict the CMEs from flaring ARs. However, owing to the moderate level of skill scores and correlation with CME productivity (Figure~\ref{Fig_CMEkin}(a)), the $SGPIL_M$ combined with the measures of the coronal magnetic field configuration would give better CME predicting capabilities. These conclusions should be tested with many more events before the relationship could be said to be robust enough to have better CME prediction capabilities. Obviously, the next step would be considering many more events and run a machine learning algorithm. The machine Learning algorithm can learn from input data and improve from experience, without human intervention. For our type of study, we can use non-linear classification machine learning algorithm, like support vector machine algorithm used in \citet{Bobra2015}, to have more quantitative rigour and check for robustness of these results.

\acknowledgments The data have been used here courtesy of NASA/SDO and HMI science team. We thank the HMI science team for the open data policy of processed vector magnetograms. N.V is a CSIR-SRF, gratefully acknowledges the funding from CSIR. P.V.R is supported by an INSPIRE grant under AORC scheme of Department of Science and Technology. We thank both referee  and statistician at ApJ for their encouraging comments and suggestions.

\bibliographystyle{apj} 

\begin{thebibliography}{68}
	\expandafter\ifx\csname natexlab\endcsname\relax\def\natexlab#1{#1}\fi
	
	\bibitem[{Abramenko(2005)}]{Abramenko2005}
	Abramenko, V.~I. 2005, \apj, 629, 1141
	
	\bibitem[{Abramenko {et~al.}(1996)Abramenko, Wang, \&
		Yurchishin}]{Abramenko1996}
	Abramenko, V.~I., Wang, T., \& Yurchishin, V.~B. 1996, \solphys, 168, 75
	
	\bibitem[{{Alissandrakis}(1981)}]{Alissandrakis1981}
	{Alissandrakis}, C.~E. 1981, \aap, 100, 197
	
	\bibitem[{{Ambastha} {et~al.}(1993){Ambastha}, {Hagyard}, \&
		{West}}]{Ambastha1993}
	{Ambastha}, A., {Hagyard}, M.~J., \& {West}, E.~A. 1993, \solphys, 148, 277
	
	\bibitem[{{Bao} {et~al.}(2002){Bao}, {Sakurai}, \& {Suematsu}}]{bao2002}
	{Bao}, S.~D., {Sakurai}, T., \& {Suematsu}, Y. 2002, \apj, 573, 445
	
	\bibitem[{{Bloomfield} {et~al.}(2012){Bloomfield}, {Higgins}, {McAteer}, \&
		{Gallagher}}]{Bloomfield2012}
	{Bloomfield}, D.~S., {Higgins}, P.~A., {McAteer}, R.~T.~J., \& {Gallagher},
	P.~T. 2012, \apjl, 747, L41
	
	\bibitem[{{Bobra} \& {Couvidat}(2015)}]{Bobra2015}
	{Bobra}, M.~G., \& {Couvidat}, S. 2015, \apj, 798, 135
	
	\bibitem[{Bokenkamp(2007)}]{Bokenkamp2007}
	Bokenkamp, N. 2007, { Statistical relationships between solar active region
		photospheric magnetic properties and flare events, PhD thesis, Stanford
		University}
	
	\bibitem[{{Calabretta} \& {Greisen}(2002)}]{Calabretta2002}
	{Calabretta}, M.~R., \& {Greisen}, E.~W. 2002, \aap, 395, 1077
	
	\bibitem[{{Chandrasekhar}(1961)}]{Chandrasekhar1961}
	{Chandrasekhar}, S. 1961, {Hydrodynamic and hydromagnetic stability}
	
	\bibitem[{{Chen} {et~al.}(2011){Chen}, {Wang}, {Shen}, {Ye}, {Zhang}, \&
		{Wang}}]{chen2011JGRA}
	{Chen}, C., {Wang}, Y., {Shen}, C., {Ye}, P., {Zhang}, J., \& {Wang}, S. 2011,
	Journal of Geophysical Research (Space Physics), 116, A12108
	
	\bibitem[{{Cheng} {et~al.}(2011){Cheng}, {Zhang}, {Ding}, {Guo}, \&
		{Su}}]{Cheng2011}
	{Cheng}, X., {Zhang}, J., {Ding}, M.~D., {Guo}, Y., \& {Su}, J.~T. 2011, \apj,
	732, 87
	
	\bibitem[{{Emslie} {et~al.}(2012){Emslie}, {Dennis}, {Shih}, {Chamberlin},
		{Mewaldt}, {Moore}, {Share}, {Vourlidas}, \& {Welsch}}]{Emslie2012}
	{Emslie}, A.~G., {Dennis}, B.~R., {Shih}, A.~Y., {Chamberlin}, P.~C.,
	{Mewaldt}, R.~A., {Moore}, C.~S., {Share}, G.~H., {Vourlidas}, A., \&
	{Welsch}, B.~T. 2012, \apj, 759, 71
	
	\bibitem[{{Falconer}(2001)}]{Falconer2001}
	{Falconer}, D.~A. 2001, \jgr, 106, 25185
	
	\bibitem[{{Falconer} {et~al.}(2003){Falconer}, {Moore}, \&
		{Gary}}]{Falconer2003}
	{Falconer}, D.~A., {Moore}, R.~L., \& {Gary}, G.~A. 2003, Journal of
	Geophysical Research (Space Physics), 108, 1380
	
	\bibitem[{{Falconer} {et~al.}(1997){Falconer}, {Moore}, {Porter}, {Gary}, \&
		{Shimizu}}]{Falconer1997}
	{Falconer}, D.~A., {Moore}, R.~L., {Porter}, J.~G., {Gary}, G.~A., \&
	{Shimizu}, T. 1997, \apj, 482, 519
	
	\bibitem[{{Gary}(1989)}]{Gary1989}
	{Gary}, G.~A. 1989, \apjs, 69, 323
	
	\bibitem[{Georgoulis \& Rust(2007)}]{Georgoulis2007}
	Georgoulis, M.~K., \& Rust, D.~M. 2007, \apjl, 661, L109
	
	\bibitem[{{Gopalswamy} {et~al.}(2009){Gopalswamy}, {Yashiro}, {Michalek},
		{Stenborg}, {Vourlidas}, {Freeland}, \& {Howard}}]{Gopalswamy2009}
	{Gopalswamy}, N., {Yashiro}, S., {Michalek}, G., {Stenborg}, G., {Vourlidas},
	A., {Freeland}, S., \& {Howard}, R. 2009, Earth Moon and Planets, 104, 295
	
	\bibitem[{{Guerra} {et~al.}(2018){Guerra}, {Park}, {Gallagher}, {Kontogiannis},
		{Georgoulis}, \& {Bloomfield}}]{Guerra2018}
	{Guerra}, J.~A., {Park}, S.-H., {Gallagher}, P.~T., {Kontogiannis}, I.,
	{Georgoulis}, M.~K., \& {Bloomfield}, D.~S. 2018, \solphys, 293, 9
	
	\bibitem[{{Guo} {et~al.}(2007){Guo}, {Zhang}, \& {Chumak}}]{Guo2007}
	{Guo}, J., {Zhang}, H.~Q., \& {Chumak}, O.~V. 2007, \aap, 462, 1121
	
	\bibitem[{{Hagino} \& {Sakurai}(2004)}]{Hagino2004}
	{Hagino}, M., \& {Sakurai}, T. 2004, \pasj, 56, 831
	
	\bibitem[{{Hagyard}(1988)}]{Hagyard1988}
	{Hagyard}, M.~J. 1988, \solphys, 115, 107
	
	\bibitem[{{Hagyard} \& {Rabin}(1986)}]{Hagyard1986}
	{Hagyard}, M.~J., \& {Rabin}, D.~M. 1986, Advances in Space Research, 6, 7
	
	\bibitem[{{Hagyard} {et~al.}(1990){Hagyard}, {Venkatakrishnan}, \&
		{Smith}}]{Hagyard1990}
	{Hagyard}, M.~J., {Venkatakrishnan}, P., \& {Smith}, Jr., J.~B. 1990, \apjs,
	73, 159
	
	\bibitem[{{Hoeksema} {et~al.}(2014){Hoeksema}, {Liu}, {Hayashi}, {Sun},
		{Schou}, {Couvidat}, {Norton}, {Bobra}, {Centeno}, {Leka}, {Barnes}, \&
		{Turmon}}]{Hoeksema2014}
	{Hoeksema}, J.~T., {Liu}, Y., {Hayashi}, K., {Sun}, X., {Schou}, J.,
	{Couvidat}, S., {Norton}, A., {Bobra}, M., {Centeno}, R., {Leka}, K.~D.,
	{Barnes}, G., \& {Turmon}, M. 2014, \solphys, 289, 3483
	
	\bibitem[{{Jing} {et~al.}(2006){Jing}, {Song}, {Abramenko}, {Tan}, \&
		{Wang}}]{Jing2006}
	{Jing}, J., {Song}, H., {Abramenko}, V., {Tan}, C., \& {Wang}, H. 2006, \apj,
	644, 1273
	
	\bibitem[{{Jing} {et~al.}(2010){Jing}, {Tan}, {Yuan}, {Wang}, {Wiegelmann},
		{Xu}, \& {Wang}}]{Jing2010}
	{Jing}, J., {Tan}, C., {Yuan}, Y., {Wang}, B., {Wiegelmann}, T., {Xu}, Y., \&
	{Wang}, H. 2010, \apj, 713, 440
	
	\bibitem[{{Kazachenko} {et~al.}(2017){Kazachenko}, {Lynch}, {Welsch}, \&
		{Sun}}]{Kazachenko2017}
	{Kazachenko}, M.~D., {Lynch}, B.~J., {Welsch}, B.~T., \& {Sun}, X. 2017, \apj,
	845, 49
	
	\bibitem[{{Kosovichev} \& {Zharkova}(2001)}]{Kosovichev2001}
	{Kosovichev}, A.~G., \& {Zharkova}, V.~V. 2001, \apjl, 550, L105
	
	\bibitem[{{Leka} \& {Barnes}(2003{\natexlab{a}})}]{Leka2003a}
	{Leka}, K.~D., \& {Barnes}, G. 2003{\natexlab{a}}, \apj, 595, 1277
	
	\bibitem[{{Leka} \& {Barnes}(2003{\natexlab{b}})}]{Leka2003b}
	---. 2003{\natexlab{b}}, \apj, 595, 1296
	
	\bibitem[{Leka \& Barnes(2007)}]{Leka2007}
	Leka, K.~D., \& Barnes, G. 2007, \apj, 656, 1173
	
	\bibitem[{Leka {et~al.}(1993)Leka, Canfield, McClymont, de~La~Beaujardiere,
		Fan, \& Tang}]{Leka1993}
	Leka, K.~D., Canfield, R.~C., McClymont, A.~N., de~La~Beaujardiere, J.-F., Fan,
	Y., \& Tang, F. 1993, \apj, 411, 370
	
	\bibitem[{{Leka} {et~al.}(1996){Leka}, {Canfield}, {McClymont}, \& {van
			Driel-Gesztelyi}}]{Leka1996}
	{Leka}, K.~D., {Canfield}, R.~C., {McClymont}, A.~N., \& {van Driel-Gesztelyi},
	L. 1996, \apj, 462, 547
	
	\bibitem[{{Liu}(2008)}]{Liu2008}
	{Liu}, Y. 2008, \apjl, 679, L151
	
	\bibitem[{{Low}(1982)}]{Low1982}
	{Low}, B.~C. 1982, \solphys, 77, 43
	
	\bibitem[{{Mason} \& {Hoeksema}(2010)}]{Mason2010}
	{Mason}, J.~P., \& {Hoeksema}, J.~T. 2010, \apj, 723, 634
	
	\bibitem[{Metcalf {et~al.}(2005)Metcalf, Leka, \& Mickey}]{Metcalf2005a}
	Metcalf, T.~R., Leka, K.~D., \& Mickey, D.~L. 2005, \apjl, 623, L53
	
	\bibitem[{{Molodensky}(1974)}]{Molodensky1974}
	{Molodensky}, M.~M. 1974, \solphys, 39, 393
	
	\bibitem[{{Olmedo} \& {Zhang}(2010)}]{Olmedo2010}
	{Olmedo}, O., \& {Zhang}, J. 2010, \apj, 718, 433
	
	\bibitem[{Pevtsov {et~al.}(1994)Pevtsov, Canfield, \& Metcalf}]{Pevtsov1994}
	Pevtsov, A.~A., Canfield, R.~C., \& Metcalf, T.~R. 1994, \apjl, 425, L117
	
	\bibitem[{{Pevtsov} {et~al.}(1995){Pevtsov}, {Canfield}, \&
		{Metcalf}}]{Pevtsov1995}
	{Pevtsov}, A.~A., {Canfield}, R.~C., \& {Metcalf}, T.~R. 1995, \apjl, 440, L109
	
	\bibitem[{{Sadykov} \& {Kosovichev}(2017)}]{Sadykov2017}
	{Sadykov}, V.~M., \& {Kosovichev}, A.~G. 2017, \apj, 849, 148
	
	\bibitem[{{Schou} {et~al.}(2012){Schou}, {Scherrer}, {Bush}, {Wachter},
		{Couvidat}, {Rabello-Soares}, {Hoeksema}, {Liu}, {Duvall}, {Akin},
		{Allard}, {Miles}, {Rairden}, {Shine}, {Tarbell}, {Title}, {Wolfson},
		{Elmore}, {Norton}, \& {Tomczyk}}]{Schou2012}
	{Schou}, J., {Scherrer}, P.~H., {Bush}, R.~I., {Wachter}, R., \& {et al}. 2012, \solphys, 275, 229
	
	\bibitem[{{Schrijver}(2007)}]{Schrijver2007}
	{Schrijver}, C.~J. 2007, \apjl, 655, L117
	
	\bibitem[{{Schrijver} {et~al.}(2005){Schrijver}, {De Rosa}, {Title}, \&
		{Metcalf}}]{Schrijver2005}
	{Schrijver}, C.~J., {De Rosa}, M.~L., {Title}, A.~M., \& {Metcalf}, T.~R. 2005,
	\apj, 628, 501
	
	\bibitem[{{Sharykin} {et~al.}(2016){Sharykin}, {Sadykov}, {Kosovichev},
		{Vargas-Dominguez}, \& {Zimovets}}]{Sharykin2016}
	{Sharykin}, I.~N., {Sadykov}, V.~M., {Kosovichev}, A.~G., {Vargas-Dominguez},
	S., \& {Zimovets}, I.~V. 2016, ArXiv e-prints
	
	\bibitem[{{Song} {et~al.}(2009){Song}, {Tan}, {Jing}, {Wang}, {Yurchyshyn}, \&
		{Abramenko}}]{Song2009}
	{Song}, H., {Tan}, C., {Jing}, J., {Wang}, H., {Yurchyshyn}, V., \&
	{Abramenko}, V. 2009, \solphys, 254, 101
	
	\bibitem[{{Song} {et~al.}(2006){Song}, {Yurchyshyn}, {Yang}, {Tan}, {Chen}, \&
		{Wang}}]{Song2006}
	{Song}, H., {Yurchyshyn}, V., {Yang}, G., {Tan}, C., {Chen}, W., \& {Wang}, H.
	2006, \solphys, 238, 141
	
	\bibitem[{{Su} {et~al.}(2014){Su}, {Jing}, {Wang}, {Wiegelmann}, \&
		{Wang}}]{Su2014}
	{Su}, J.~T., {Jing}, J., {Wang}, S., {Wiegelmann}, T., \& {Wang}, H.~M. 2014,
	\apj, 788, 150
	
	\bibitem[{{Sun}(2013)}]{Sun2013}
	{Sun}, X. 2013, ArXiv e-prints
	
	\bibitem[{{Sun} {et~al.}(2015){Sun}, {Bobra}, {Hoeksema}, {Liu}, {Li}, {Shen},
		{Couvidat}, {Norton}, \& {Fisher}}]{Sun2015}
	{Sun}, X., {Bobra}, M.~G., {Hoeksema}, J.~T., {Liu}, Y., {Li}, Y., {Shen}, C.,
	{Couvidat}, S., {Norton}, A.~A., \& {Fisher}, G.~H. 2015, \apjl, 804, L28
	
	\bibitem[{{Tian} {et~al.}(2002){Tian}, {Wang}, \& {Wu}}]{Tian2002}
	{Tian}, L., {Wang}, J., \& {Wu}, D. 2002, \solphys, 209, 375
	
	\bibitem[{{Tiwari} {et~al.}(2009){Tiwari}, {Venkatakrishnan}, \&
		{Sankarasubramanian}}]{Tiwari2009}
	{Tiwari}, S.~K., {Venkatakrishnan}, P., \& {Sankarasubramanian}, K. 2009,
	\apjl, 702, L133
	
	\bibitem[{{Toriumi} {et~al.}(2017){Toriumi}, {Schrijver}, {Harra}, {Hudson}, \&
		{Nagashima}}]{Toriumi2017}
	{Toriumi}, S., {Schrijver}, C.~J., {Harra}, L.~K., {Hudson}, H., \&
	{Nagashima}, K. 2017, \apj, 834, 56
	
	\bibitem[{{T{\"o}r{\"o}k} \& {Kliem}(2005)}]{Torok2005}
	{T{\"o}r{\"o}k}, T., \& {Kliem}, B. 2005, \apjl, 630, L97
	
	\bibitem[{{Vemareddy}(2017)}]{Vemareddy2017b}
	{Vemareddy}, P. 2017, \apj, 845, 59
	
	\bibitem[{{Vemareddy} {et~al.}(2012){Vemareddy}, {Ambastha}, \&
		{Maurya}}]{Vemareddy2012}
	{Vemareddy}, P., {Ambastha}, A., \& {Maurya}, R.~A. 2012, \apj, 761, 60
	
	\bibitem[{{Vemareddy} {et~al.}(2015){Vemareddy}, {Venkatakrishnan}, \&
		{Karthikreddy}}]{Vemareddy2015}
	{Vemareddy}, P., {Venkatakrishnan}, P., \& {Karthikreddy}, S. 2015, Research in
	Astronomy and Astrophysics, 15, 1547
	
	\bibitem[{{Vemareddy} \& {Zhang}(2014)}]{Vemareddy2014}
	{Vemareddy}, P., \& {Zhang}, J. 2014, \apj, 797, 80
	
	\bibitem[{{Wang} {et~al.}(1994{\natexlab{a}}){Wang}, {Ewell}, {Zirin}, \&
		{Ai}}]{Wang1994}
	{Wang}, H., {Ewell}, Jr., M.~W., {Zirin}, H., \& {Ai}, G. 1994{\natexlab{a}},
	\apj, 424, 436
	
	\bibitem[{{Wang} {et~al.}(2006){Wang}, {Song}, {Jing}, {Yurchyshyn}, {Deng},
		{Zhang}, {Falconer}, \& {Li}}]{WangHM2006}
	{Wang}, H.-M., {Song}, H., {Jing}, J., {Yurchyshyn}, V., {Deng}, Y.-Y.,
	{Zhang}, H.-Q., {Falconer}, D., \& {Li}, J. 2006, \cjaa, 6, 477
	
	\bibitem[{{Wang} {et~al.}(1996){Wang}, {Shi}, {Wang}, \& {Lue}}]{WangJ1996}
	{Wang}, J., {Shi}, Z., {Wang}, H., \& {Lue}, Y. 1996, \apj, 456, 861
	
	\bibitem[{{Wang} {et~al.}(1994{\natexlab{b}}){Wang}, {Xu}, \&
		{Zhang}}]{WangT1994}
	{Wang}, T., {Xu}, A., \& {Zhang}, H. 1994{\natexlab{b}}, \solphys, 155, 99
	
	\bibitem[{{Wang} \& {Zhang}(2007)}]{YumingWang2007}
	{Wang}, Y., \& {Zhang}, J. 2007, \apj, 665, 1428
	
	\bibitem[{Zhang \& Bao(1999)}]{ZhangHQ1999}
	Zhang, H., \& Bao, S. 1999, \apj, 519, 876
	
	\bibitem[{Zirin \& Wang(1993)}]{Zirin1993a}
	Zirin, H., \& Wang, H. 1993, \solphys, 144, 37
	
\end{thebibliography}

%
%

\end{document}